\documentclass[%
 reprint,
 amsmath,amssymb,
 aps,
 pra,showkeys,
]{revtex4-2}
\usepackage[T1]{fontenc}
\usepackage{graphicx}% Include figure files
\usepackage{dcolumn}% Align table columns on decimal point
\usepackage{bm}% bold math
\usepackage{subcaption} % for subfigure environment
\usepackage{amsthm}
\usepackage{braket}
\usepackage{hyperref}% add hypertext capabilities
\newtheorem{theorem}{Theorem}
\newtheorem{lemma}{Lemma}
\newtheorem{corollary}{Corollary}
\theoremstyle{remark}
\newtheorem{remark}{Remark}
\DeclareUnicodeCharacter{0229}{\k{e}}

\begin{document}

\preprint{APS/2311.10433}

\title{Task Scheduling Optimization with Direct Constraints from a Tensor Network Perspective}% Force line breaks with \\

\author{Alejandro Mata Ali}
 \email{alejandro.mata.ali@gmail.com}
\affiliation{Instituto Tecnológico de Castilla y León, Burgos, Spain}
\author{Iñigo Perez Delgado}
\email{iperezde@ayesa.com}
\affiliation{i3B Ibermatica, Parque Tecnológico de Bizkaia, Ibaizabal Bidea, Edif. 501-A, 48160 Derio, Spain}
\author{Beatriz García Markaida}
\email{b.garcia@ibermatica.com}
\affiliation{i3B Ibermatica, Parque Tecnológico de Bizkaia, Ibaizabal Bidea, Edif. 501-A, 48160 Derio, Spain}
\author{Aitor Moreno Fdez. de Leceta}
\email{aitormoreno@lksnext.com}
\affiliation{Quantum Technologies and Systems Unit, LKS Next, MONDRAGON Corporation, Goiru 7, 20500 Arrasate-Mondragón, Gipuzkoa, Spain}

\date{\today}% It is always \today, today,

\begin{abstract}
This work presents a novel method for task optimization in industrial plants using quantum-inspired tensor network technology. This method obtains the best possible combination of tasks on a set of machines with directed constraints while minimizing the total execution cost. With this method, an exact and explicit solution of the problem is provided. This algorithm constructs a tensor network representation of the tensor which provides the solution of the problem. This method is improved in order to reduce the computational complexity of the solution computation, using problem preprocessing, new techniques of condensation of logical constraints, optimization of the value determination technique with previously calculated results, reuse of intermediate computations, and iterative relations for constraints. Three algorithms for computation are presented: the main algorithm, the iterative algorithm which adds only the minimal amount of necessary constraints, and the genetic algorithm which combines the iterative algorithm with basic genetic algorithms. Finally, a simple version of both algorithms was implemented, and their performance was tested, all publicly available.
\end{abstract}

%Use showkeys class option if keyword
\maketitle

%\tableofcontents

\section{Introduction}
The distribution of production resources is widely considered one of the most interesting and useful problems in industry \cite{FlowShop1, FlowShop2, FlowShop3}. This distribution problem can be stated as a constraint combinatorial problem, and the time required to obtain the exact optimal solution usually scales exponentially with the size of the problem. However, in real-world applications usually it is not required to obtain the optimal solution, only a good enough solution computed in a reasonable time is needed. There exist some classical heuristic approximations which are capable to reduce the computational time so much, which has led them to be considered great applied methods, such as \textit{genetic algorithms} \cite{Genetic} or \textit{particle swarm optimization} \cite{Enjambre}. Even with these methods, obtaining a compatible and optimal solution may be extremely costly for certain types of problem.

Quantum computing applied to industrial cases has become very interesting because of its computational power. Some of the best known and most promising quantum algorithms for combinatorial optimization are the \textit{Quantum Approximate Optimization Algorithm} (QAOA) \cite{QAOA}, \textit{Variational Quantum Eigensolver} (VQE) \cite{VQE}, and \textit{Quantum Annealing for Constrained Optimization} (QACO) \cite{Annealing}.  However, these algorithms are limited due to the current \textit{Noisy intermediate-scale quantum} (NISQ) state of small quantum computers with notable noise.
 
Due to this, great expectations have arisen with quantum-inspired methods, which are based on imitating certain quantum processes in classical systems to improve their performance. An example is \textit{digital annealing} to solve \textit{Quadratic Unconstrained Binary Optimization} (QUBO) problems \cite{Digital}. Another branch inspired by quantum theory is that of \textit{tensor networks} \cite{Tensor}, a classical technology based on the use of linear algebra that allows to classically simulate quantum systems both exact and approximately and compress information efficiently. Several algorithms are available in tensor networks to address various combinatorial optimization problems \cite{TTOpt,GEO,Combin}. However, it is desirable to have specialized algorithms for particular cases in order to reduce their computational complexity and memory requirements as much as possible and improve their performance.

A specific optimization problem for industrial processes is to assign tasks to be performed to a set of machines, given a set of constraints on the tasks that can be performed by one machine depending on the task performed by a different machine. This case is interesting because both the execution times of the tasks and the constraints between them can be extracted from a historical record of the corresponding manufacturing plant, without having to logically deduce them or having to perform tests.

This paper presents an equation that exactly solves the problem and two algorithms to compute it efficiently in an approximate way. It is called the \textit{Directed Constraint Task Scheduler Tensor Network Solver} (DCTSTN). In this approach,the quantum-inspired tensor networks formalism \textit{MeLoCoToN}~\cite{melocoton} is combined with iterative methods and genetic algorithms. According to the available information, this is the first work to apply MeLoCoToN to constraint optimization problems and perform a case of \textit{Motion Onion}, described in the same MeLoCoToN work. The main novelty contributions of this work are as follows:
\begin{itemize}
    \item An exact and explicit tensor network-based equation that solves the minimum-total-execution-cost problem with a set of directed constraints.
    \item An iterative approximated algorithm to compute the solution given by the equation, which in the limit is an exact algorithm.
    \item A genetic algorithm to compute approximate solutions to the problem.
    \item Performance analysis with an available Python implementation.
\end{itemize}

This work is structured as follows. First, Sect.~\ref{sec: description} describes the problem to be addressed and a brief background of the state-of-the-art approach to solving similar problems. Second, Sec.~\ref{sec: algorithm} introduces the algorithm to efficiently obtain tensor network that provides the solution equation for the problem. Third, Sec.~\ref{sec: improvement} describes several improvements in the construction of the tensor network to reduce the amount of resources needed for its computation. Then, Sec.~\ref{sec: generalization} describes how to generalize the algorithm for more general problems. Next, Sec.~\ref{sec: computation} describes the different algorithms to compute the solution from the tensor network construction. Finally, in Sec.~\ref{sec: experiments} the performance of the algorithms is tested with several instances and sizes.

All required code is publicly available on the GitHub repository \href{https://github.com/DOKOS-TAYOS/Task_Scheduler_with_Tensor_Networks}{https://github.com/DOKOS-TAYOS/Task\_Scheduler\_with\_Tensor\_Networks} and a Streamlit application at \href{https://task-scheduler-with-tensor-networks.streamlit.app/}{https://task-scheduler-with-tensor-networks.streamlit.app/}.

\section{Description of the problem}\label{sec: description}

\begin{figure}[h]
    \centering
    \includegraphics[width=\linewidth]{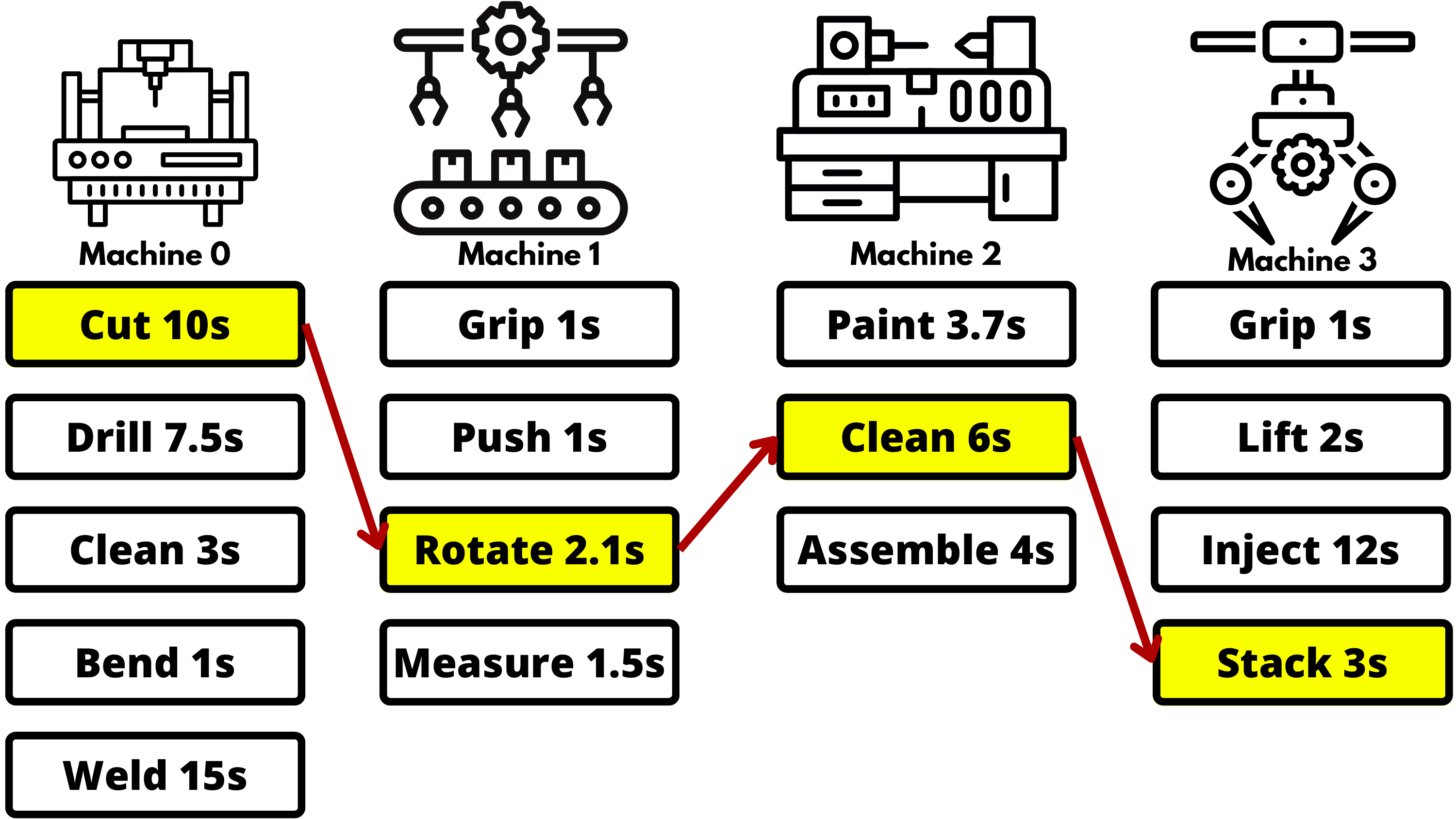}
    \caption{Example of problem instance with $m=4$ machines and $P=\lbrace5,4,3,4\rbrace$ tasks, with solution $\vec{x}=(0,2,1,3)$.}
    \label{fig:Machines}
\end{figure}

The problem to solve is the optimal distribution of tasks in a set of machines with some constraints on multiple sets of tasks. That is, the problem has a set of $m$ machines and on $i$-th machine there are $P_i$ possible tasks, with an execution time $T_{ij}$ for the $j$-th task on $i$-th machine, with $i\in [0, m-1]$ and $j \in [0, P_i-1]$. It also has a set of directed constraints for these task combinations. An example of constraints would be: ``If machine 0 performs task 2 and machine 1 performs task 4, machine 2 must perform task 3.''. Fig.~\ref{fig:Machines} shows an example of a problem instance, where each machine must do one task, and they are correlated by restrictions.

The optimal solution is a combination of machines that satisfies the constraints with the lowest total execution cost. This is essential to increase the productivity of industrial plants. Obviously, other indicators can be used, such as cost instead of execution time. It is assumed that there are no extra execution times due to the order of task execution or moving from one cycle to another. However, the method will be extended in Ssec. \ref{sec: generalization}.

The combination of tasks is expressed as a vector $\vec{x}$ such that each element of the vector is the task executed on the machine of the corresponding position. Then,
$\vec{x}=(1,3,2,3)$, indicates that machine 0 is assigned task 1, machine 1 is assigned task 3, machine 2 is assigned task 2, and machine 3 is assigned task 3. In other words, $x_i$ is the task executed in the $i$-th machine.

As with any optimization problem, a cost function is needed. In this case, this function is the total execution cost, which in the present setting is the sum of the individual execution times. This can be written as a Tensor quadratic unconstrained discrete optimization~\cite{QUDO} problem, with a sum of the individual local costs  
\begin{equation}
C(\vec{x})=\sum_{i=0}^{m-1} T_{i,x_i}.
\end{equation}
In this work, the objective function is therefore additive, namely $C(\vec{x})$, rather than the classical parallel-machine makespan $C_{\max}$. For this reason, the optimized quantity is referred to throughout as the total execution cost.
The set of all constraints is obtained by combining all the individual constraints. Each constraint is denoted as $R^k$, for $k\in [0, n_{r}-1]$, which is $n_{r}$ the number of constraints.
The constraints are written as a list with two elements such that the first one is conditional and the second one conditioned. That is, for the above example, ``if machine 0 has task 2 and machine 1 has task 4, then machine 2 must have task 3.'', the constraint string would be: $R^0 = [ [2,4,\ ''], [2,3] ]$, where $''$ implies that this machine does not condition the other ones, but it can be the conditioned one.

This problem can be viewed as a version of other scheduling problems such as the Resource-Constrained Project Scheduling Problem (RCPSP)~\cite{bottleneck_RCPS,extension_RCPSP}, Flow Shop Scheduling Problem~\cite{FlowShop1,FlowShop2,FlowShop3} or Job Shop Scheduling Problem~\cite{Mixed_JobShop}. This type of problem is generally known as NP problems~\cite{NP_Complete,Precedence,Resource_complex,Complexity_lands}. However, because of its importance in industry, there are several methods to solve them in an approximate way. First, genetic algorithms have been widely applied to scheduling problems~\cite{FlowShop1,Genetic,genetic_rcps,genetic_flexible,genetic_job_shop,genetic_flow_shop}, but with a limitation on parameterization of the problems and their size. Another line of solving this kind of problems is the integer programming~\cite{Mixed_JobShop,integer_flow}, with limitations in scalability, requiring a large number of variables, and difficulty in the implementation of exotic constraints. The tabu search has also been studied to solve this kind of problem~\cite{tabu_search_job_shop, Tabu_job_shop_2} and iterative local search~\cite{iterative_flow_shop,iterative_job_shop}, which works in certain contexts, without over-structured instances. Quantum solvers have also been used, such as quantum annealing~\cite{Q_annealing_rcps,Q_annealing_flexible,Q_Annealer_JSSP_Dwave,q_annealer_flexible_2} or QAOA~\cite{qaoa_jssp}, with the limitation of current quantum hardware.

\section{Tensor network Equation}\label{sec: algorithm}
The core of the solution equation is inspired by the simulation of a quantum system with qudits by means of a tensor network, taking advantage of the non-unitary operations that the latter allows. However, it is important to recall that this is still a classical formalism, so an explanation from both the classical and quantum computation points of view will be presented. The deduction method relies on improving the algorithm of~\cite{Combin} for the implementation of restrictions to reduce complexity and memory cost, combining it with the tensors presented in~\cite{QUDO} for the minimization part, and obtaining the final result with the MeLoCoToN~\cite{melocoton} techniques. This means that this equation is a particular case of MeLoCoToN and the first developed for constraint optimization problems.

To check all combinations of tasks, it is necessary to create a tensor $\psi$, with all elements $\psi_{\vec{x}}$ associated with the possible solution $\vec{x}$. If its elements have a value that decreases with $C(\vec{x})$, and incompatible combinations have zero elements, the problem is reduced to searching for the position of the maximum element of the tensor. It is possible to efficiently construct the tensor network representation of this tensor, and then extract its maximum element position. This tensor network with the extraction method provides the equation that solves the problem.

From a quantum point of view, the construction consists of the following:
\begin{enumerate}
    \item The creation of the initial uniform superposition state in the qudits: $\ket{\psi^0}=\sum_{\vec{x}} \ket{\vec{x}}$.
    \item The application of an imaginary time evolution, so that the amplitude of a combination depends on its cost and the damping constant $\tau$: $\ket{\psi^1}=\sum_{\vec{x}} e^{-\tau C(\vec{x})}\ket{\vec{x}}$.
    \item The removal of states by applying constraints by means of projectors: $\ket{\psi^2}=\sum_{\vec{x}} R^0R^1\dots R^{n_r-1}e^{-\tau C(\vec{x})}\ket{\vec{x}}$.
    \item The measurement and extraction of the basis state with the maximum amplitude of the superposition.
\end{enumerate}

From a classical point of view, it consists of the following:
\begin{enumerate}
    \item The creation of the initial represented tensor where all elements are equal, representing all possible combinations: $\psi^0_{x_0,x_1,\dots}=1$.
    \item The application of an operation to assign to every possible combination element a value proportional to its cost, with a damping constant $\tau$: \mbox{$\psi^1_{x_0,x_1,\dots}=e^{-\tau C(\vec{x})}$}.
    \item The removal of elements of incompatible combinations by means of projective operations: \mbox{$\psi^2_{x_0,x_1,\dots}= \mathcal{R}^0_{x_0,x_1,\dots}\mathcal{R}^1_{x_0,x_1,\dots}\dots \mathcal{R}^{n_r-1}_{x_0,x_1,\dots}e^{-\tau C(\vec{x})}$}.
    \item The measurement and extraction of the maximum element position of the represented tensor.
\end{enumerate}

This section is structured as follows. First, Ssec.~\ref{ssec: initial} presents the initialization and evolution tensor layers. Second, Ssec.~\ref{ssec: layer constr} presents the creation of the layers of tensors that implement the constraints, and Ssec.~\ref{ssec: set} presents the application of the contraint layers. Then, Ssec.~\ref{ssec: measure} presents the value extraction protocol. Finally, Ssec.~\ref{ssec: equation} presents the exact and explicit equation from the tensor network.

\subsection{Initial system and imaginary time evolution}\label{ssec: initial}
\begin{figure}
    \centering
    \includegraphics[width=0.75\linewidth]{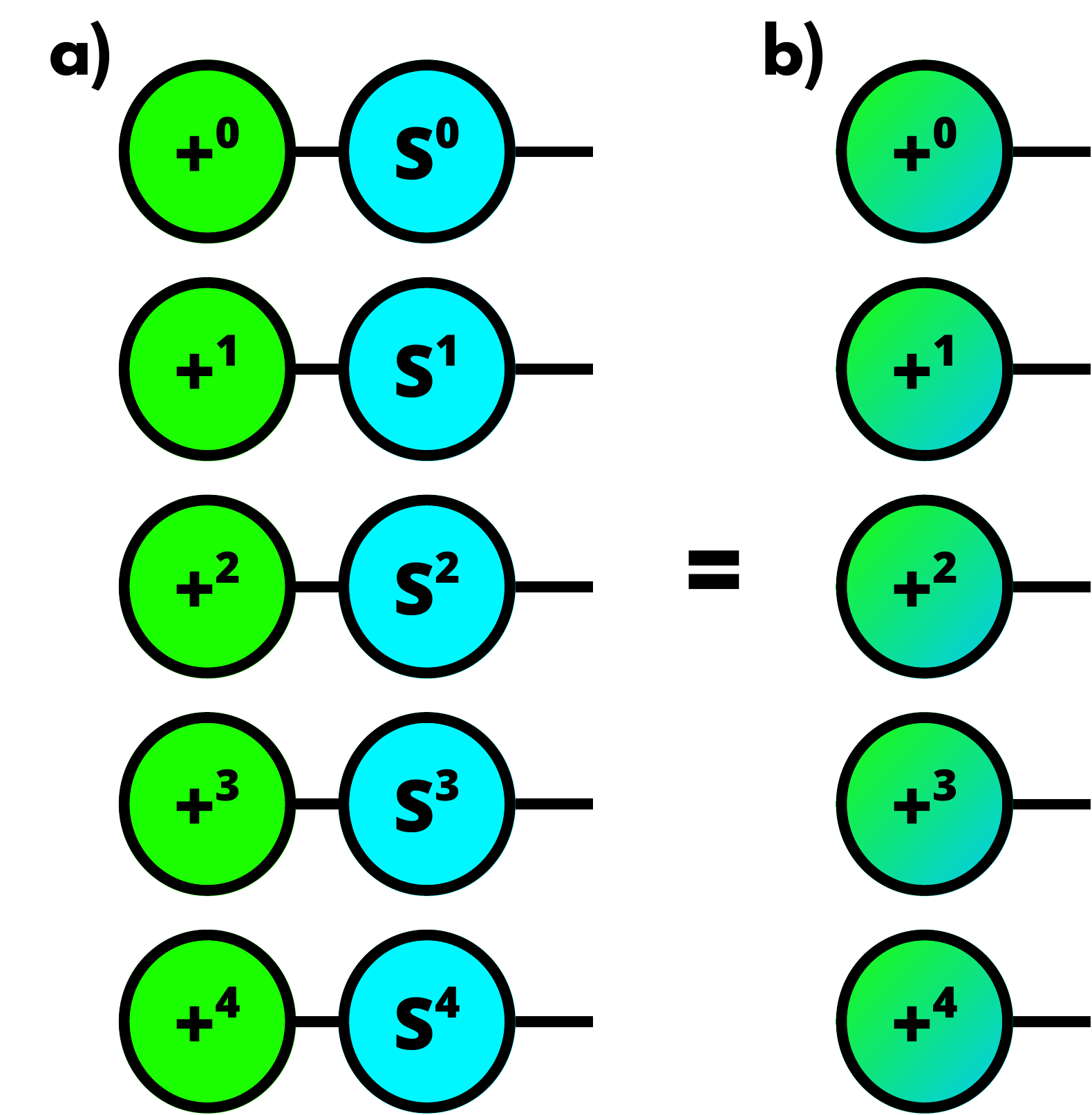}
    \caption{Initialization nodes. a) Initial superposition layer `+' and evolution layer $S$. b) New condensated initialization layer `+'.}
    \label{fig: initialization}
\end{figure}

The algorithm starts by generating the tensor that considers all possible solutions and assigns a value to them. This is performed in two steps. First, a set of $m$ tensors `+' is created, where the $i$-th tensor represents the $i$-th index of the represented tensor $\psi^0$, associated with the variable $x_i$. These tensors have all their elements equal to one, and the $i$-th tensor has $P_i$ elements, each representing each possible value for the variable $x_i$. The tensor product of all these tensors is equal to
\begin{equation}
    \psi^0_{\vec{x}} = +^0_{x_0}+^1_{x_1}\cdots +^{m-1}_{x_{m-1}} = 1.
\end{equation}
Then, an imaginary time evolution operator is applied to each tensor, obtaining the tensor
\begin{equation}
    \psi^1_{\vec{x}} = e^{-\tau C(\vec{x})}.
\end{equation}
This ensures that the combinations with lower cost have larger elements than the combinations with higher cost. Increasing $\tau$ increases the separation between different costs, while for finite $\tau$ the separation required for extraction depends on the cost gaps of the instance, as discussed in Sec.~\ref{ssec: measure} and Sec.~\ref{sec: experiments}. So, this tensor performs the unconstrained optimization part of the problem. To do this evolution, the algorithm uses the first two layers of the paper method \cite{QUDO}, adapted to this problem cost function by removing the bond indexes in the evolution layer. This layer is shown in Fig.~\ref{fig: initialization} a.

In quantum terms, it makes a superposition of all possible tasks and applies the imaginary time evolution. The state after these steps is
\begin{equation}
    |\psi\rangle = \sum_{\vec{x}} e^{-\tau C(\vec{x})} |\vec{x}\rangle,
\end{equation}
so that the higher the cost of the combination, the lower its amplitude exponentially.

However, since the costs depend only on the variable and not on its neighbors, the `$S$' layer of imaginary time evolution can be absorbed in the `+' layer of initialization itself, as shown in Fig.~\ref{fig: initialization} b. The resulting represented tensor is
\begin{equation}
    \psi^1_{\vec{x}} = \prod_{i=0}^{m-1}e^{-\tau T_{i,x_i}} = e^{-\tau C(\vec{x})},
\end{equation}
being the new `+' tensors
\begin{equation}
    +^i_{j} = e^{-\tau T_{i,j}}.
\end{equation}

In quantum terms, the minimization operation can also be expressed as
\begin{align}
    |\psi^1\rangle =& \sum_{x_0} e^{-\tau T_{0,x_0}} \ket{x_0}\otimes\sum_{x_1} e^{-\tau T_{1,x_1}} \ket{x_1}\otimes\dots\nonumber\\
    &\dots\otimes \sum_{x_{m-1}} e^{-\tau T_{m-1,x_{m-1}}} \ket{x_{m-1}} =\nonumber\\
    =&|+(T)\rangle^0\otimes |+(T)\rangle^1\otimes\dots\otimes|+(T)\rangle^{m-1},
\end{align}
so it can be obtained by means of the tensor product of the $m-1$ vectors already minimized locally;
\begin{equation}
    |+(T)\rangle^i = \sum_{x_i} e^{-\tau T_{i,x_i}} \ket{x_i}.
\end{equation}

If a task has to be deleted, its corresponding evolution element value has to be replaced with a 0, equivalent to setting its cost to infinity.

\subsection{Constraint layer creation}\label{ssec: layer constr}
In order to impose the constraint on the tensor, a set $\lbrace \mathcal{R}^r\rbrace$ of tensor layers is required. This construction provides a similar output to that presented in~\cite{cons_train}. Each layer represents a tensor that projects the $\psi$ into the space that satisfies that constraint. That is, the $r$-th constraint layer sets to zero the values of the elements corresponding to the combinations that do not satisfy the $r$-th constraint.

The constraint layers can be build as a Matrix Product Operator (MPO), with one tensor connected to each index of the $\psi$ tensor. Each tensor of the MPO checks the value of its corresponding variable and sends it to the next tensor if its value corresponds to the constraint. Finally, the tensor connected with the constrained variable projects the value of the variable into the imposed by the constraint if the other variables have the values that activate it. In other words, if the constraint is $R^0 = [ [2,4,\ ''], [2,3] ]$, the first two tensors of the layer check if the values of its variables are $2$ and $4$, respectively, and if this is the case, the third tensor multiplies by zero the values associated to values different from $3$ in the final variable. In this case, it sets the components $\psi_{2,4,0}, \psi_{2,4,1}, \psi_{2,4,2}, \psi_{2,4,4}$ to zero and $\psi_{2,4,3}$ multiplied by one. This construction requires only sending through the tensors if the first part of the constraint is satisfied, so it can be performed with an MPO of bond dimension equal to $2$.

The required tensors have five types, inspired by the Grover oracle circuit~\cite{Grover}: Ctrl, Cctrl, cProj, CcProj, and Id. The Ctrl tensors simply send to the next tensor if the value of its variable is required for the activation of the constraint. The Cctrl tensors perform the same task, but only if they receive the signal that the previous variables have the correct values. The cProj tensors perform the projection only if they receive that the previous variables have the correct values. The CcProj performs the same, but only if they receive the signal by both sides. The Id tensor only passes the signal that it receives.

\begin{figure}[h]
    \centering
    \includegraphics[width=\linewidth]{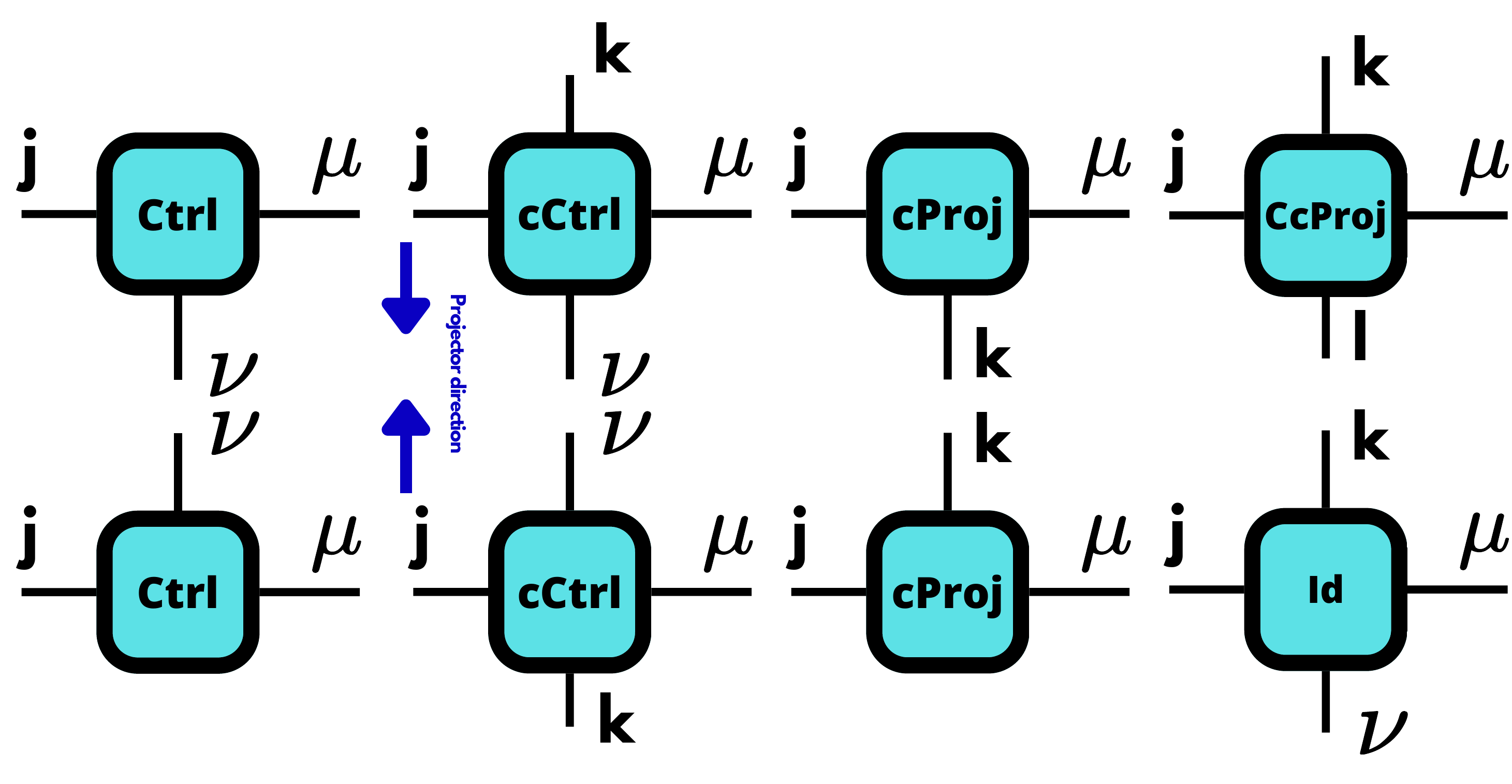}
    \caption{Name of the indexes for the different tensors depending on the relative position of the projector tensor.}
    \label{fig:Types}
\end{figure}

Following the notation presented in~\cite{melocoton} for sparse logical tensors and the index names in Fig.~\ref{fig:Types}, the tensors are defined in the following way.
\begin{itemize}
    \item $Ctrl^{i,a}$: Ctrl tensor in the $i$-th variable index, for a constraint with the first part $x_i=a$. Passes $1$ by its vertical index if $x_i= a$ and $0$ otherwise.
    \begin{equation}
    \begin{gathered}
        \mu = j,\quad \nu = \delta_{j,a},\\
        Ctrl^{i,a}_{j\mu \nu} = 1,
    \end{gathered}
    \end{equation}
    being $\delta_{j,a}$ the Kronecker delta, which equals $1$ if $j=a$ and $0$ otherwise.
    \item $Cctrl^{i,a}$: Cctrl tensor in the $i$-th variable index, for a constraint with the first part $x_i=a$. Passes $1$ by its vertical index if $x_i=a$ and receives $1$ from the previous tensor, and $0$ otherwise.
    \begin{equation}
    \begin{gathered}
        \mu = j,\quad \nu = \delta_{j,a}k,\\
        Cctrl^{i,a}_{jk\mu \nu} = 1.
    \end{gathered}
    \end{equation}
    \item $cProj^{i,a}$: cProj tensor in the $i$-th variable index, for a constraint with the second part $x_i=a$ if the first part is satisfied. Only passes $a$ by its horizontal index if it receives $1$ from the previous tensor, and all values otherwise.
    \begin{equation}
    \begin{gathered}
        \mu = j,\\
        cProj^{i,a}_{jk\mu} = (1-k)+k\delta_{j,a}.
    \end{gathered}
    \end{equation}
    \item $CcProj^{i,a}$: CcProj tensor in the $i$-th variable index, for a constraint with the second part $x_i=a$ if the first part is satisfied. Only passes $a$ by its horizontal index if it receives $1$ from the previous tensor and the next tensor, and all values otherwise.
    \begin{equation}
    \begin{gathered}
        \mu = j,\\
        CcProj^{i,a}_{jkl\mu} = (1-kl)+kl\delta_{j,a}.
    \end{gathered}
    \end{equation}
    \item $Id^i$: Id tensor in the $i$-th variable index. Passes the values of the vertical and horizontal indexes without changes of conditions. It is used for variables that are not included in the constraint.
    \begin{equation}
    \begin{gathered}
        \mu = j,\nu = k,\\
        Id^{i}_{jk\mu\nu} = 1.
    \end{gathered}
    \end{equation}
\end{itemize}

\begin{figure}[h]
    \centering
    \includegraphics[width=\linewidth]{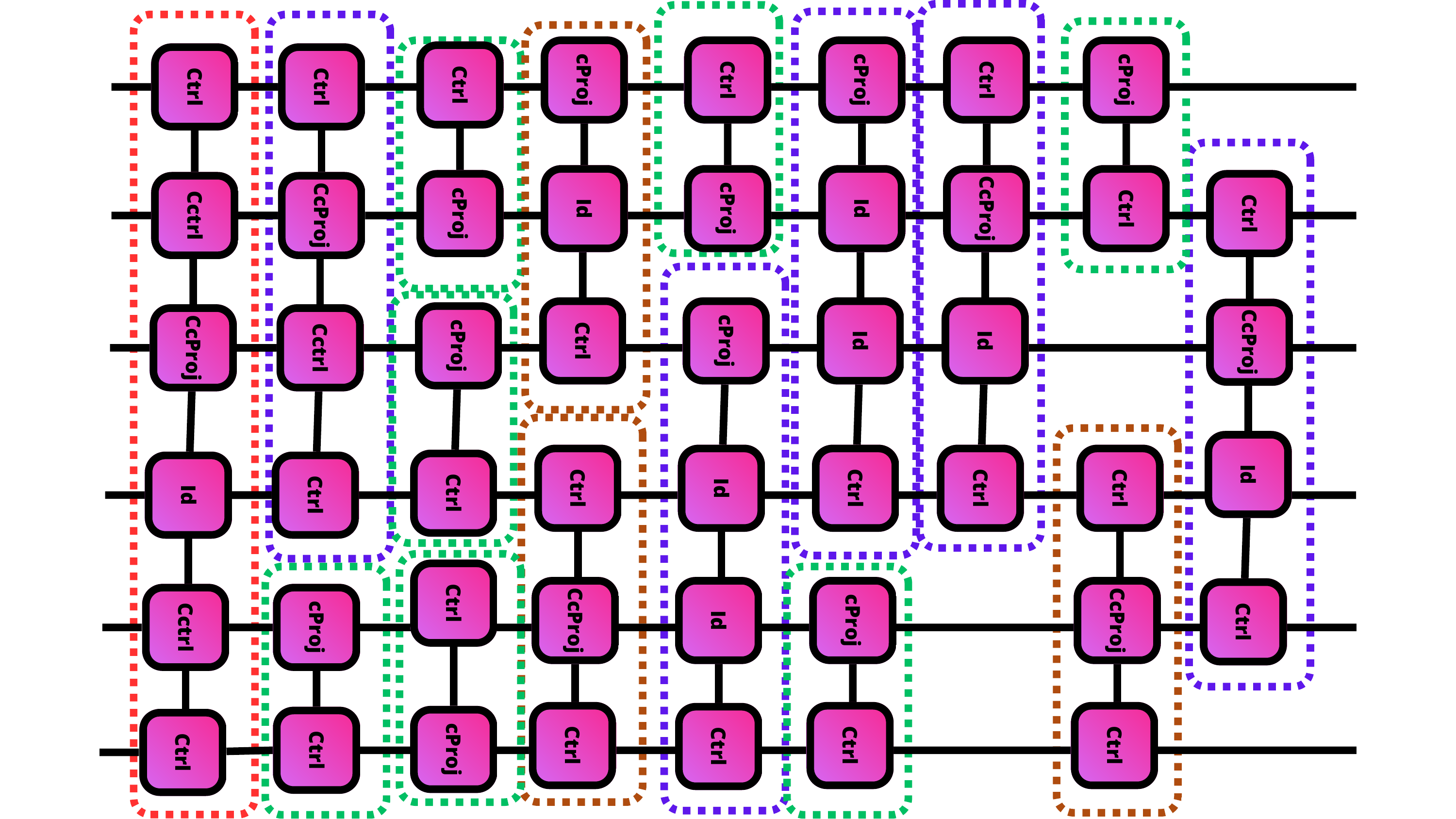}
    \caption{Example of constraint layers connected in a compact way. The represented constraints are $[[a,a,'','',a,a],[2,a]]$, $[[a,'',a,a,'',''],[1,a]]$, $[['','','','','',a],[4,a]]$, $[[a,a,'','','',''],[1,a]]$, $[['','',a,a,'',''],[2,a]]$, $[['','','','',a,a],[5,a]]$, etc.}
    \label{fig:constraint Layers}
\end{figure}

The constraint layer is constructed by starting the MPO with the Ctrl or cProj tensor at the position of the first variable in the rule, if it is in the first or second part, respectively. Then, include a Cctrl tensor for each variable in the first part of the constraint, except the last one included if it is after the conditioned variable. Include cProj or CcProj in the conditioned variable position if it is after the last variable of the first part or before. Finally, if the last variable of the constraint is in the first part, include a Ctrl tensor in its position. The other variables with are not in the constraint and are between the variables that are in the constraint have an Id tensor in its positions. Each Ctrl, Cctrl, and cProj have their indexes oriented in a consistent way to give the signals to the projector and the projector receive them. This process is repeated for each constraint, obtaining layers as the ones represented in Fig.~\ref{fig:constraint Layers}. Each $r$-th constraint layer represents the tensor $\mathcal{R}^r$, with elements $\mathcal{R}^r_{\vec{y}^r,\mathcal{S}^r(\vec{x})}$, being $\mathcal{S}^r(\vec{x})$ the subset of variable indices that this constraint layer needs. Due to the construction of these layers, for the element of the represented tensor to be non-zero, it should satisfy that $\vec{y}^r=\mathcal{S}^r(\vec{x})$, and the $\mathcal{S}^r(\vec{x})$ values satisfy the $r$-th constraint.

\subsection{Constraint operator creation}\label{ssec: set}
With the constraints layers set, they are joined one after another, as in Fig. \ref{fig:constraint Layers}, and finally applied to the $\psi^1$ tensor. It is important that when connecting the constraint layers, the tensor network needs to be as compact as possible to improve its computation. That is, there should be as few intermediate gaps as possible. This will help the contraction algorithms reduce the memory and time costs. The represented global constraint operator is
\begin{equation}
    \mathcal{R} = \prod_{r=0}^{n_r-1} \mathcal{R}^r.
\end{equation}

Applying these tensor layers, the $\psi$ tensor now is
\begin{equation}
    \psi^2_{\vec{x}} = \prod_{r=0}^{n_r-1} \mathcal{R}^r_{\mathcal{S}^r(\vec{x}),\mathcal{S}^r(\vec{x})}e^{-\tau C(\vec{x})}.
\end{equation}
This tensor only has non-zero elements in the compatible combinations, and each element value is exponentially lower if its corresponding cost is larger. So, the position of the larger element of this tensor is th

In quantum terms, the current quantum state is a superposition in which each remaining element satisfies the constraints. This state is
\begin{equation}
    |\psi(C, R, \tau)\rangle = \sum_{\vec{x}} \mathcal{R}\ e^{-\tau C(\vec{x})} \ket{\vec{x}}.
    \label{eq: state minimum}
\end{equation}

\subsection{Variable value extraction}
\label{ssec: measure}
The final step is to extract the position of the largest element of the represented tensor, as this is the one with the lowest cost that satisfies the constraints. It can be done by means of \textit{Half partial traces}~\cite{QUDO}. This is based on the fact that the system has a peak value, that is, an element with a value sufficiently larger than the others. Therefore, when a Half partial trace is performed, the maximum at each index should be the same as the global one. This is performed summing over all indices, but the one of the variable to be extracted. If there is a value in the tensor that is large enough, one sum would be larger than the others, and this sum would be the one that includes this element. So, this allows to determine which value of that index contains the desired element, because it is equal to the value of the position of the largest element over this summation.

\begin{figure}[h]
    \centering
    \includegraphics[width=\linewidth]{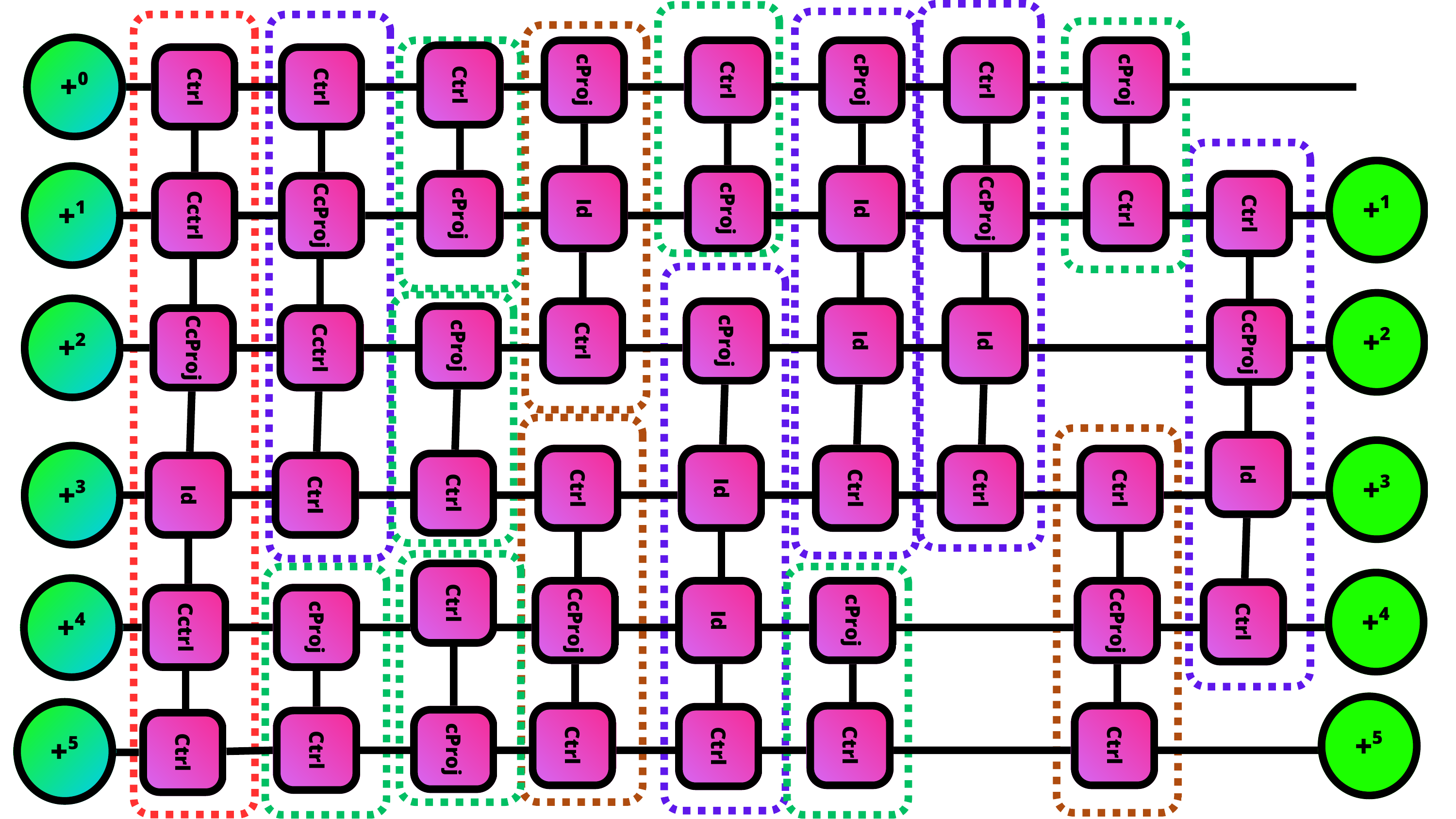}
    \caption{Tensor network performing the optimization, constraint and traces for determine the first variable.}
    \label{fig:Measure}
\end{figure}

This is done by connecting a set of nodes with all elements equal to one at the end of the last constraint layer, except for the index corresponding to the variable to be checked, as shown in Fig.~\ref{fig:Measure}. This leads to
\begin{equation}
    \psi^3_{x_0} = \sum_{x_1, x_2, \dots}\psi^2_{x_0, x_1, x_2,\dots}
\end{equation}

For any machine $i$ and task value $a$, the corresponding extracted component can be written as
\begin{equation}
    \omega^i_a(\tau)=\sum_{\vec{x}\in\mathcal{F}:x_i=a} e^{-\tau C(\vec{x})},
\end{equation}
where $\mathcal{F}\subseteq\mathcal{X}$ denotes the set of compatible assignments and $\mathcal{X}=\prod_{i=0}^{m-1}\{0,\dots,P_i-1\}$ is the finite assignment space.

\begin{theorem}[Exact recovery from marginal extraction]
    \label{thm:marginal-extraction}
    Let $\mathcal{F}\subseteq\mathcal{X}$ be the feasible set induced by all directed constraints. Assume that $\mathcal{F}\neq\emptyset$ and that there exists a unique minimizer
    \begin{equation}
        \vec{X} = \arg\min_{\vec{x}\in\mathcal{F}} C(\vec{x}).
    \end{equation}
    For $\tau>0$, the tensor network represents
    \begin{equation}
        \psi_{\tau}(\vec{x}) = e^{-\tau C(\vec{x})}\mathbf{1}_{\mathcal{F}}(\vec{x}).
    \end{equation}
    Then, for every machine $i$ and task value $a\in[0,P_i)$,
    \begin{equation}
        \lim_{\tau\rightarrow\infty}\frac{\omega^i_a(\tau)}{\sum_{b=0}^{P_i-1}\omega^i_b(\tau)}=\delta_{a,X_i}.
    \end{equation}
    Consequently,
    \begin{equation}
        X_i = \lim_{\tau\rightarrow\infty}\arg\max_{a\in[0,P_i)}\omega^i_a(\tau).
    \end{equation}
\end{theorem}
\begin{proof}
    Let $C^*=C(\vec{X})$ and
    \begin{equation}
        Z(\tau)=\sum_{\vec{x}\in\mathcal{F}} e^{-\tau C(\vec{x})}.
    \end{equation}
    Since $\mathcal{F}$ is finite and $\vec{X}$ is the unique minimizer,
    \begin{equation}
        Z(\tau)=e^{-\tau C^*}\left[1+\sum_{\vec{x}\in\mathcal{F}\setminus\{\vec{X}\}} e^{-\tau(C(\vec{x})-C^*)}\right].
    \end{equation}
    For every $\vec{x}\in\mathcal{F}\setminus\{\vec{X}\}$ one has $C(\vec{x})-C^*>0$, so all suboptimal terms vanish as $\tau\rightarrow\infty$. Hence $e^{\tau C^*}Z(\tau)\rightarrow 1$.

    If $a=X_i$, then
    \begin{equation}
        \omega^i_{X_i}(\tau)=e^{-\tau C^*}+\sum_{\substack{\vec{x}\in\mathcal{F}\setminus\{\vec{X}\}\\x_i=X_i}} e^{-\tau C(\vec{x})},
    \end{equation}
    so $e^{\tau C^*}\omega^i_{X_i}(\tau)\rightarrow 1$. If $a\neq X_i$, every term contributing to $\omega^i_a(\tau)$ has cost strictly larger than $C^*$, and therefore $e^{\tau C^*}\omega^i_a(\tau)\rightarrow 0$. Dividing by $Z(\tau)=\sum_b\omega^i_b(\tau)$ gives the stated limit. The argmax statement follows because, for sufficiently large $\tau$, the unique largest marginal component is the one indexed by $X_i$.
\end{proof}
\begin{remark}
    If $\mathcal{F}=\emptyset$, the final tensor is identically zero and the normalization above is not defined. If the optimum is degenerate, the marginals may exhibit ties, and extracting each coordinate independently need not select a coherent optimal assignment. The degenerate case is discussed in Sec.~\ref{sec: generalization}.
\end{remark}

In the resulting vector, the position of the element whose absolute value is the largest is the correct value of $x_0$, so the task of the first machine. The same procedure can be performed to determine the other variables, changing only the final layer. To determine the $i$-th variable, the initialization and constraint layers are connected as before. Now, the vector of ones are connected to each free index, but the one of the $i$-th variable, the one to be determined.

\subsection{Exact and Explicit Equation}\label{ssec: equation}
With the construction of the tensor network, each variable can be determined by argmax to the result of the contraction of the tensor network. $\vec{\omega}^i(\tau)$ is defined as the tensor network created to determine the $i$-th variable, that is, the generalization of the $\psi^3_{x_0}$ construction to an arbitrary variable index. Each tensor network is independent of the previous one and depends only on the problem and the parameter $\tau$, so it does not depend on the solution. By Theorem~\ref{thm:marginal-extraction}, this leads to the result
\begin{equation}
    X_i = \lim_{\tau\rightarrow\infty} \arg \max_{a\in[0,P_i)} \omega^i_a(\tau).
\end{equation}
Taking into account that a tensor network is an equation, this is also an equation. However, to obtain a more explicit function, the binary expression of each variable can be used. This means that instead of having one variable $X_i$ with $P_i$ possible values, there are $\log_2 P_i$ binary variables $X_{i,k}$. This can be performed by splitting the final free index of $\vec{\omega}^i(\tau)$ into a set of binary indices and performing the Half partial trace with them to determine every binary component of the $X_i$ variable. If a variable has $P_i$ that is not a power of two, it is extended with new tasks with infinite cost. Let $b_{i,k}(a)\in\{0,1\}$ be the $k$-th bit of the task value $a$ at machine $i$. The final contraction with the $(-1,1)$ vector is then
\begin{equation}
    \Omega_{i,k}(\tau)=\sum_{\vec{x}\in\mathcal{F}} \left(2b_{i,k}(x_i)-1\right)e^{-\tau C(\vec{x})}.
\end{equation}
Under the hypotheses of Theorem~\ref{thm:marginal-extraction}, $\Omega_{i,k}(\tau)$ has the sign of the optimal bit for sufficiently large $\tau$, and therefore
\begin{equation}
    X_{i,k} = \lim_{\tau\rightarrow\infty} H(\Omega_{i,k}(\tau)),
\end{equation}
where $H(\cdot)$ is the Heaviside step-function.

\section{Improvements for real computation}\label{sec: improvement}
The contraction of this tensor network can be made with an algorithm that looks for the most efficient route of contraction or by first contracting all the tensors of the last machine, then this resulting tensor with all those of the previous one, and so on until the whole network is completed. The latter scheme will result, taking into account that all $P_k=P$ for simplicity, in a computational complexity of $O(2^{n_r}n_rP^2m)$ per variable in the worst case. Even with reuse of intermediate computations as described in~\cite{QUDO}, the global complexity remains $O(2^{n_r} n_rP^2 m)$. This will depend significantly on the particular case to be solved and the structure of the resulting layers for the constraints, as well as their ordering, so it could have significantly lower complexity.

In order to reduce the number of tensors to use, the number of operations, and the amount of memory to use, several steps of compression of the constraints are applied to the tensor network to be contracted. First, Ssec.~\ref{ssec: preproc} presents a pre-processing step to prepare the problem for optimizing the execution of the resolution algorithm. Second, Ssec.~\ref{ssec: constraints} presents all the process to condensate the constraints and optimize them for its application for an efficient further contraction. Then, Ssec.~\ref{ssec: reduction extraction} describes an improved way to extract the values of the variables taking advantage of the previous variables values. 

\subsection{Pre-processing}\label{ssec: preproc}
Before the construction of the tensor network equation, the problem has to be pre-processed. The first step is to normalize the costs. The intention is for all possible costs to be in the range $(-1,1)$. Thus, the time evolution maintains the represented tensor elements in the range $(e^{-\tau},e^{\tau})$, avoiding overflows or too small gaps. This helps the computation algorithm not have to adjust $\tau$ for each problem manually. This can be done by rescaling all the times of the problem with the sum of the maximum times and the sum of the minimum times for each machine.

The next step is to sort the machines so that the tensor network has the lowest possible number of tensors. This will be better understood in Ssec. \ref{ssec: constraints}. To do this, the machines that appear in more constraints are placed closer together in the most central area of the problem. This means that if machine 0 appears in 3 constraints, machine 1 appears in 2 constraints, machine 2 in 1 constraint, machine 3 in 7 constraints, and machine 4 in 3 constraints, the machines will be sorted in the order $(2, 4, 3, 0, 1)$, changing, for example, the state $\vec{x}=(4, 5, 10, 2, 7)$ to $\vec{x}'=(10, 7, 2, 4, 5 )$. The reason for doing this is that machines that appear more in the constraints will have more connections in the tensor network with other machines than those that appear less. In this way, it reduces the number of tensors by eliminating many Id tensors whose only purpose is to connect two distant machines. From now on, all states, times, and constraints are already organized and normalized.

\subsection{Constraints} \label{ssec: constraints}
In this subsection, it is explained how to apply the constraints to the system. First, a rearrangement and condensation of the constraints are performed. The condensation is done to reduce the memory scaling with the number of constraints, and the rearrangement is done to improve the condensation. These two steps are performed to obtain a single layer of bond dimension $d+1$ that agglomerates $d$ constraints instead of having $d$ layers of one constraint each, which would have a final bond dimension of $2^d$. In this way, the bond dimension required can be exponentially reduced in a simple way.

\subsubsection{Constraint grouping}\label{sssec: group}
First, the constraints are grouped into sets that start and end on the same machines and have the same conditioned machine, as this is a condensation requirement. These sets are the constraints that can be condensed in the next step. In case the conditioned machine is before the first conditioning machine, all the constraints that have the same final conditioning machine and that conditioned machine can be joined to the set of the constraints that have the same first conditioning machine. This is the \textit{initial extreme case}. Similarly, in the case the conditioned machine is after the last conditioning machine, all the constraints that have the same initial conditioning machine and that conditioned machine can be joined to the set with the same last conditioning machine. This is the \textit{final extreme case}.

\subsubsection{Constraint condensation}\label{sssec: condens}
Within each group, the condensation begins with the generation of subgroups of up to $P_{e_G}$ constraints, where $e_G$ denotes the first conditioning machine in the standard and initial extreme cases and the last conditioning machine in the final extreme case. The conditioning task on that machine is never repeated within the subgroup. This is done to avoid summing in the next step the compatible states differently according to their extremal relation to the constraint. There are actually better constraint compression schemes, using complex techniques, but they are so complicated to generalize that they are left for a possible future study. 

Let $G$ be one of these condensed subgroups, containing $d$ constraints. By construction, if the $r$-th and $s$-th constraints of $G$ are different, then the required values at that machine are different, namely $a_{r,e_G}\neq a_{s,e_G}$ for $r\neq s$.

\begin{lemma}[Single-label invariant in a condensed subgroup]
    \label{lem:single-label}
    For every assignment $\vec{x}$ and every bond position of the condensed MPO associated with $G$, the set of still-active candidate constraints has cardinality at most one. Therefore the bond state can be encoded by a single label in $\{0,1,\dots,d\}$, where $0$ means that no candidate constraint remains active and $r+1$ means that the unique active candidate is the $r$-th constraint of the subgroup. In particular, the condensed MPO has bond dimension $d+1$.
\end{lemma}
\begin{proof}
    At the extreme conditioning machine $e_G$, the injectivity of $r\mapsto a_{r,e_G}$ implies that the observed value $x_{e_G}$ can match at most one constraint of the subgroup. Hence, immediately after that first test, there is at most one active candidate. Every subsequent tensor only checks additional literals of the same candidate: if one of them fails, the propagated label is reset to $0$; otherwise, the same non-zero label is preserved. No tensor can branch a non-zero label into two different non-zero labels. Therefore the invariant ``at most one active candidate'' is preserved all along the chain, and a scalar signal in $\{0,1,\dots,d\}$ is sufficient. In the uniform case this gives $d\leq P_{e_G}=P$, while in the non-uniform case it gives $d\leq P_{e_G}$.
\end{proof}

\subsubsection{Creating constraint layers} \label{sssec: layers}
In order to impose such restrictions, a set of constraint layers is created. However, in this case, the signal between them indicates which constraint has been activated. If the first Ctrl of the layer finds that its associated machine is in the task that indicates the third constraint, it will tell the following tensor to verify if its machine is in the task associated with the third constraint. Then, the Ctrl tensor will send the next the signal $3$ through its vertical index. If it is not in the value of none of the constraints, the returned signal is zero. If Cctrl receives the signal $3$ and its variable is at the corresponding value given by the third constraint, it will return $3$ to the following tensor. Otherwise, it will return zero because no constraint is activated. And so on until reaching the cProj, which in case of receiving that all the conditioning machines are in the values of the third constraint, it will force its machine to be in the task of the machine conditioned in this constraint, by means of a projection. The CcProj has controls on both sides, so it has to receive the same signal on both sides to project. If it receives 0 or different values on both sides, the constraints conditions will not have been met, and therefore the projection will not apply.

Tensors are described as:

\begin{itemize}
    \item $Ctrl^{i,\vec{a}}$: Ctrl tensor in the $i$-th variable index, for a set of constraints with the first part $x_i=a_r$ for the $r$-th constraint. Passes $r+1$ by its vertical index if $x_i= a_r$ and $0$ otherwise.
    \begin{equation}
    \begin{gathered}
        \mu = j,\quad \nu = \sum_r (r+1)\delta_{j,a_r},\\
        Ctrl^{i,\vec{a}}_{j\mu \nu} = 1.
    \end{gathered}
    \end{equation}
    \item $Cctrl^{i,\vec{a}}$: Cctrl tensor in the $i$-th variable index, for a set of constraints with the first part $x_i=a_r$ for the $r$-th constraint. Passes $r+1$ by its vertical index if $x_i= a_r$ and receives $r+1$ from the previous tensor, and $0$ otherwise.
    \begin{equation}
    \begin{gathered}
        \mu = j,\quad \nu = \sum_r (r+1)\delta_{j,a_r}\delta_{k,r+1},\\
        Cctrl^{i,\vec{a}}_{jk\mu \nu} = 1.
    \end{gathered}
    \end{equation}
    \item $cProj^{i,\vec{a}}$: cProj tensor in the $i$-th variable index, for a set of constraints with the second part $x_i=a_r$ for the $r$-th constraint if the first part is satisfied. Only passes $a_r$ by its horizontal index if it receives $r+1$ from the previous tensor, and all values otherwise.
    \begin{equation}
    \begin{gathered}
        \mu = j,\\
        cProj^{i,\vec{a}}_{jk\mu} = \delta_{k,0}+\sum_r\delta_{k,r+1}\delta_{j,a_r}.
    \end{gathered}
    \end{equation}
    \item $CcProj^{i,\vec{a}}$: CcProj tensor in the $i$-th variable index, for a set of constraints with the second part $x_i=a_r$ for the $r$-th constraint if the first part is satisfied. Only passes $a_r$ by its horizontal index if it receives $r+1$ from the previous tensor and the next tensor, and all values otherwise.
    \begin{equation}
    \begin{gathered}
        \mu = j,\\
        CcProj^{i,\vec{a}}_{jkl\mu} =
        \begin{cases}
            1, & k=0 \text{ or } l=0 \text{ or } k\neq l,\\
            \delta_{j,a_{k-1}}, & k=l\neq 0.
        \end{cases}
    \end{gathered}
    \end{equation}
    When two non-zero incoming labels are different, they correspond to two different candidate rules selected by the left and right parts of the chain. Since no single rule has been selected by both sides, no consequent can be enforced at the target site, and therefore all target values must pass. By Lemma~\ref{lem:single-label}, each side can carry at most one active label, but the two sides may still disagree at the target site.
    \item $Id^i$: Id tensor in the $i$-th variable index. Passes the values of the vertical and horizontal indexes without changes of conditions. It is used for variables that are not included in the constraint.
    \begin{equation}
    \begin{gathered}
        \mu = j,\nu = k,\\
        Id^{i}_{jk\mu\nu} = 1.
    \end{gathered}
    \end{equation}
\end{itemize}

In the uniform case, where $P_{e_G}=P$ for every condensed subgroup, this condensation allows the tensor network to be composed of a set of $O(n_r/P)$ constraint layers, with bond dimension $O(P)$. More generally, each condensed layer has bond dimension $O(P_{e_G})$ for its corresponding extreme machine. Under the same uniform assumption, this improves the computational complexity from $O(2^{n_r} n_rP^2 m)$ to $O(P^{n_r/P} n_r m)$. These tensors can be condensed in more complex ways to improve the number of constraints joined, but this could be studied in future research.

\subsection{Extraction reduction}\label{ssec: reduction extraction}
The extraction process can be optimized taking into account that in the $i$-th variable determination step, the values of the $i-1$ previous variables are known. The part of the tensor network dedicated to the already known variables could be neglected, introducing into the tensors of the current variable the signals they would receive. Then, the same scheme is repeated again, but taking into account only the variables to be determined. To further improve the complexity of each step, the constraint layers that are not needed are removed because they are the bottleneck of the contraction algorithm. Before each determination step, the following modifications are made:
\begin{enumerate}
    \item All the constraints in which the previous machine conditioned another machine and had to have a different value than the one found are removed. This is because obviously these constraints will never be activated.
    \item If the previous machine conditions another machine and its value is the one found, the constraint is kept by eliminating the dependence on the previous machine, since it will always satisfy this part.
    \item If the only conditioner of a constraint is the previous machine and has the obtained value, this constraint is eliminated, and the task of the target machine is forced to be the one imposed by the constraint. This is because it will always be active. This allows to determine the variable without contracting the tensor network.
    \item If a constraint has the previous machine as conditional and its value is the obtained, it disappears, since it is always satisfied.
    \item If a constraint has as conditioned the previous machine with a value different from the one obtained, the constraint is replaced by one that makes the conditioned combination cannot occur. This is because if that combination is found, the constraint would not be satisfied.
\end{enumerate}
For each variable determined, either by contraction or by this process, the process is repeated for that variable, adding the new known information. That is, in addition to the above, if there is a constraint in which the determined machines are the only conditioners for another machine and they all have the values obtained, this constraint is eliminated, and the task of the other machine is forced to be the one imposed by the constraint. This is because it will always be active.

\section{Generalizations}\label{sec: generalization}

This tensor network construction can be adapted to more general problems by performing some modifications. First, if each machine has a cost related to the previous one in the chain, having a nearest-neighbor Tensor Quadratic Unconstrained Discrete Optimization (T-QUDO) cost function
\begin{equation}
C(\vec{x})=\sum_{i=0}^{m-2} T_{i,i+1,x_i,x_{i+1}}.
\end{equation}
In this case, the only change is to replace the initialization layer with the initialization and evolution layers of~\cite{QUDO}.

The second important aspect to be aware is that a certain constraint/constraint set could have additional information. The cost of a particular state can be increased by changing the filtered element in the projector in the corresponding constraint from $1$ to exponential with the extra cost and $\tau$. This adds additional terms to the cost in a simple way. Also, the target machine can have a set of tasks instead of a single task by adding more non-null elements to the projectors.

The third generalization is the degenerate case. In this case, the equation can be obtained only with the argmax, making every variable depend on the previously obtained and always choosing the first value obtained in case the argmax returns more than one value. The equation is also exact and explicit, but less elegant. However, in practical situations, it is convenient, because it allows to extract all the $M$ degenerate solutions performing the same algorithm $M$ times, but every time choosing different variable values in the cases where argmax returns more than one value.

\section{Computation algorithms}\label{sec: computation}
The computation of the solutions from the equation requires to contract a 2D grid network which scales exponentially with the number of constraints. One way to avoid excessive scaling was to condense the constraints, but it is still excessive. To avoid this scaling, two alternative approximate algorithms are proposed: the iterative algorithm in Ssec.~\ref{ssec: iterative} and the genetic algorithm in Ssec.~\ref{ssec: genetic}. These methods can be combined.

\subsection{Iterative algorithm}\label{ssec: iterative}
This method is based on not applying all the constraints at once at the beginning. Only one compressed constraint is applied, and in case of failure, the first non-satisfied compressed constraint is applied, repeating this process iteratively until all constraints are satisfied. Every additional constraint reduces the feasible region, because the feasible set associated with a larger set of constraints is the intersection of the feasible sets of all the constraints applied. The key property used by the iterative algorithm is the following certification result.

\begin{theorem}[Certification from constraint subsets]
    Let $S$ be a set of constraints and let $\hat{S}\subseteq S$. Define
    \begin{equation}
        \mathcal{F}_S=\{\vec{x}\in\mathcal{X}:\vec{x}\text{ satisfies every constraint in }S\},
    \end{equation}
    and
    \begin{equation}
        \operatorname{Opt}(S)=\arg\min_{\vec{x}\in\mathcal{F}_S} C(\vec{x}).
    \end{equation}
    If $\hat{\vec{x}}\in\operatorname{Opt}(\hat{S})$ and $\hat{\vec{x}}\in\mathcal{F}_S$, then $\hat{\vec{x}}\in\operatorname{Opt}(S)$.
\end{theorem}
\begin{proof}
    Since $\hat{S}\subseteq S$, every assignment that satisfies all constraints in $S$ also satisfies all constraints in $\hat{S}$, and therefore $\mathcal{F}_S\subseteq\mathcal{F}_{\hat{S}}$. Let $\vec{y}\in\mathcal{F}_S$. Then $\vec{y}\in\mathcal{F}_{\hat{S}}$, so the optimality of $\hat{\vec{x}}$ for the relaxed problem gives
    \begin{equation}
        C(\hat{\vec{x}})\leq C(\vec{y}).
    \end{equation}
    This is true for every $\vec{y}\in\mathcal{F}_S$. Since $\hat{\vec{x}}\in\mathcal{F}_S$ by hypothesis, it is feasible for the full problem and has cost no larger than any feasible assignment of the full problem. Hence $\hat{\vec{x}}\in\operatorname{Opt}(S)$.
\end{proof}
\begin{corollary}
    If $\operatorname{Opt}(S)=\{\vec{X}\}$ is a singleton, then any $\hat{\vec{x}}\in\operatorname{Opt}(\hat{S})$ that satisfies all constraints in $S$ must satisfy $\hat{\vec{x}}=\vec{X}$.
\end{corollary}

Therefore, if the optimum of the relaxed problem defined by a subset of constraints already satisfies all original constraints, it is certified to be globally optimal for the full problem. If the full problem has a unique optimum, the previous corollary shows that the relaxed optimum must coincide with it. This means that, with fewer constraint layers, the problem can still be solved exactly. This is important because the complexity of the contraction algorithm scales exponentially with the number of constraints, so reducing the number of active layers can substantially improve the computation. However, determining the minimal useful subset of constraints is not trivial. For this reason, the iterative approach adds constraints until the relaxed optimum satisfies all the original ones.

Then, the optimization algorithm is as follows:  
\begin{enumerate}
    \item The algorithm first considers the relaxed problem with $\hat{S}=\emptyset$. If an obtained minimizer $\hat{\vec{x}}\in\operatorname{Opt}(\emptyset)$ already belongs to $\mathcal{F}_S$, it finishes, because then $\hat{\vec{x}}\in\operatorname{Opt}(S)$. If not, it continues.
    \item The first unsatisfied condensed constraint is added to $\hat{S}$, also adding the unmatched constraints that are compatible with it in order to condense them and eliminate more combinations and iterations. The relaxed problem associated with the new subset $\hat{S}$ is then solved again. If the obtained minimizer belongs to $\mathcal{F}_S$, the algorithm finishes. Otherwise, this step is repeated until feasibility for $S$ is reached or the iteration limit is met.
    \item If it reaches the limit of iterations, that is, the maximum number of condensed constraint sets allowed to be added to $\hat{S}$, it finishes with a negative result: no solution that follows all constraints of the chosen subset has been found. 
\end{enumerate}
It is important to note that this is an algorithm for determining a solution certified by a sufficient subset of constraints, not necessarily for finding a minimum-cardinality subset, due to the extra constraints added during condensation. From this subset, a smaller certifying subset can be searched for by repeating the process without extra constraints condensation. This is the first application of the \textit{Motion Onion} method~\cite{melocoton}.

In the worst case, this algorithm requires the use of all constraints, which takes much more operations than the simple complete contraction approach. However, if the constraints are redundant and not fine-tuned to be all needed, as in problems as the traveling salesman problem, it would require a small subset of constraints. This is a relevant point for real-world problems, where its constraints are extracted from historical data or heuristics.

\subsection{Genetic algorithm }\label{ssec: genetic}
This algorithm is based on solving several reduced versions of the main problem, called its subproblems, taking advantage that their solutions are computable with more efficiency. Each subproblem considers only a subset of tasks for each machine and a subset of constraints.  With this approach, a genetic algorithm can be applied that defines every subproblem as an individual. Each individual in the population has the following attributes:  
\begin{itemize}
    \item Chromosomes: a subset of tasks that each machine can perform and a subset of constraints, consistent with the subset of tasks.  
    \item Result: solution obtained from the tensor network equation and the computation algorithm. This can also be calculated with the iterative method in cases with too many constraints.
    \item Cost: Cost of the result.
\end{itemize}

The method is as follows: 
\begin{enumerate}
    \item Random initialization of the population.  
    \item Computation of the result for each individual. Only a proportion of the best individuals are kept, using the cost function as the evaluator.  
    \item With these individuals, a crossover is performed. Crosses pairs of parents from the previous generation and corrects the associated constraints, removing the ones that will never be activated.
    \item Creation of a set of mutated individuals from the parents and correct the associated constraints. Mutations are performed by changing some tasks to perform in the machines.  
    \item Check for repeat individuals for their removal.  
    \item Addition of new randomly created individuals to make up for the missing ones, to have the number of individuals it should have.
    \item Addition of new randomly selected constraints for the new generation individuals, until they have a fixed amount. This is performed to make each individual subproblem representative with respect to the main problem.
    \item Repetition of steps $2\rightarrow 6$ until the convergence criteria are met or the fixed maximum number of generations is reached. 
\end{enumerate}

The chromosome crossover is performed by exchanging, within the same machine, the possible active tasks a number of times, chosen at random between the two individuals.
The constraint correction is applied to each individual each time it is created. If a constraint cannot be activated, it is removed from the individual. For example, if the constraint requires a task that is not included in the individual. For individuals with missing constraints in their heritage, new compatible constraints are added. The output of the algorithm is a set of possible results sorted by their quality.

\section{Experiments}\label{sec: experiments}
The algorithm is tested by creating simulated cases for the three cases of normal algorithm, iterative algorithm, and genetic algorithm with iterative algorithm. In the experiments below we use $\tau=100$, which was sufficient to separate the optimum in the tested instances. This should not be interpreted as a universal value: for finite $\tau$, the correctness of the marginal extraction depends on the relevant cost gap between the best feasible assignment and the closest suboptimal feasible assignments, as well as on the number of suboptimal states contributing to each marginal. Informally, if $\Delta$ is a lower bound on that gap and $N$ bounds the number of competing suboptimal assignments in a marginal bucket, a sufficient scale is $\tau \gtrsim \log(N)/\Delta$. The algorithms have been implemented in the TensorNetwork library~\cite{TN_lib}, and are performed on the CPU, with an Intel(R) Core(TM) i7-14700HX 2.10 GHz and 16 GB RAM. This implementation is a simplification of the techniques explained in order to analyze the core of the algorithms. It does not perform the iterative removal of constraints and variable tensors during the determination of variable values. It also uses the automatic contraction method provided for the library. 

Case generation is performed by choosing the number of machines, the number of tasks per machine, and the number of constraints. First, a list of random times is generated with a uniform distribution between $0$ and $10$ for each machine and task. Then a set of constraints is created that are compatible with each other, i.e. the same condition cannot lead to two different tasks on the same machine.

\begin{figure*}
    \centering
    \begin{subfigure}[b]{0.32\linewidth}
        \centering
        \includegraphics[width=\linewidth]{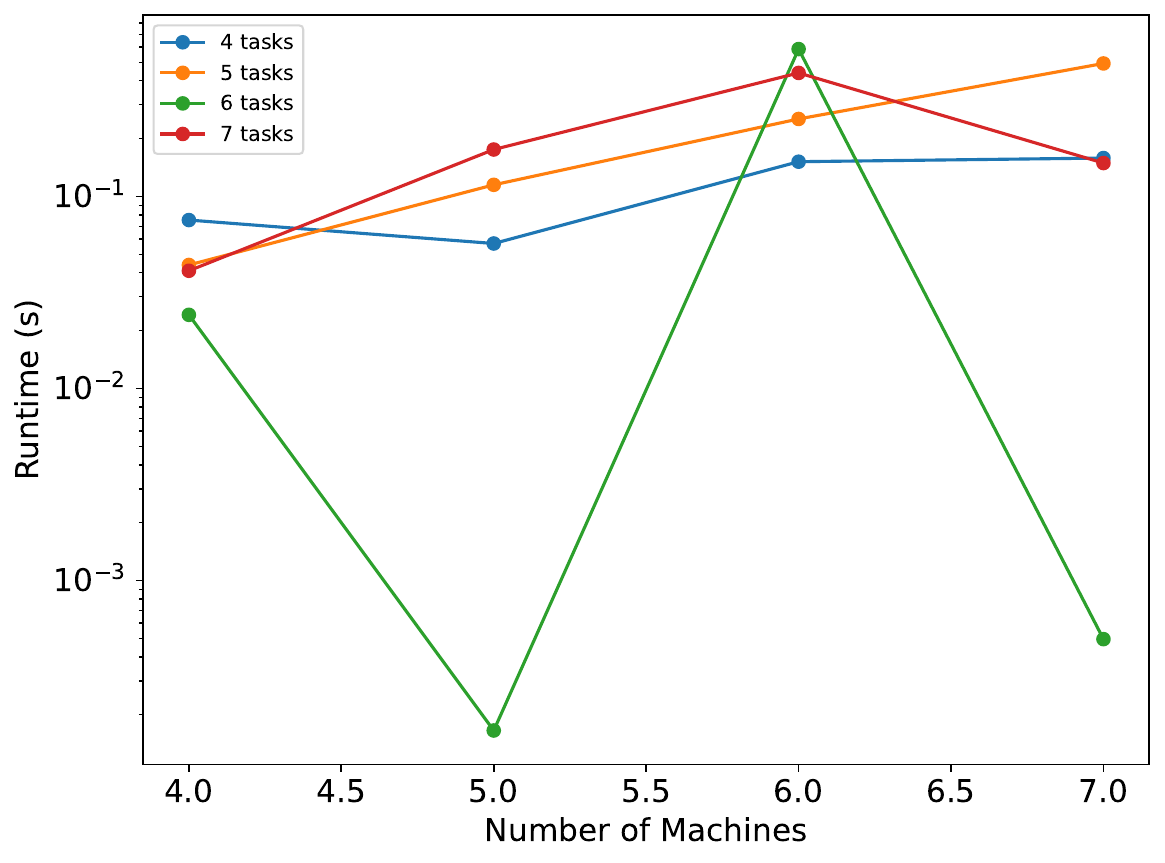}
        \caption{Runtime against number of machines $m$, for different number of tasks. There are $15$ constraints.}
        \label{fig: scaling mach base}
    \end{subfigure}
    \hfill
    \begin{subfigure}[b]{0.32\linewidth}
        \centering
        \includegraphics[width=\linewidth]{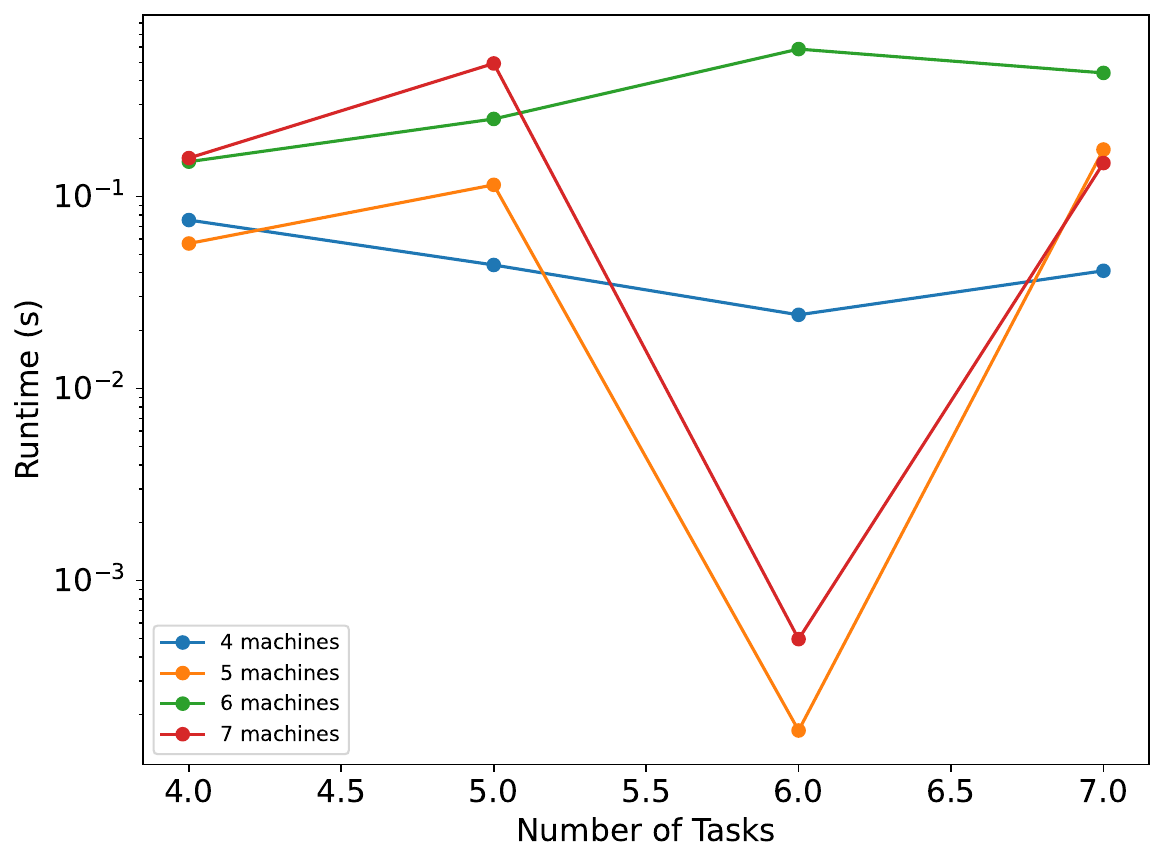}
        \caption{Runtime against number of tasks $P$, for different number of machines. There are $15$ constraints.}
        \label{fig: scaling tasks base}
    \end{subfigure}
    \hfill
    \begin{subfigure}[b]{0.32\linewidth}
        \centering
        \includegraphics[width=\linewidth]{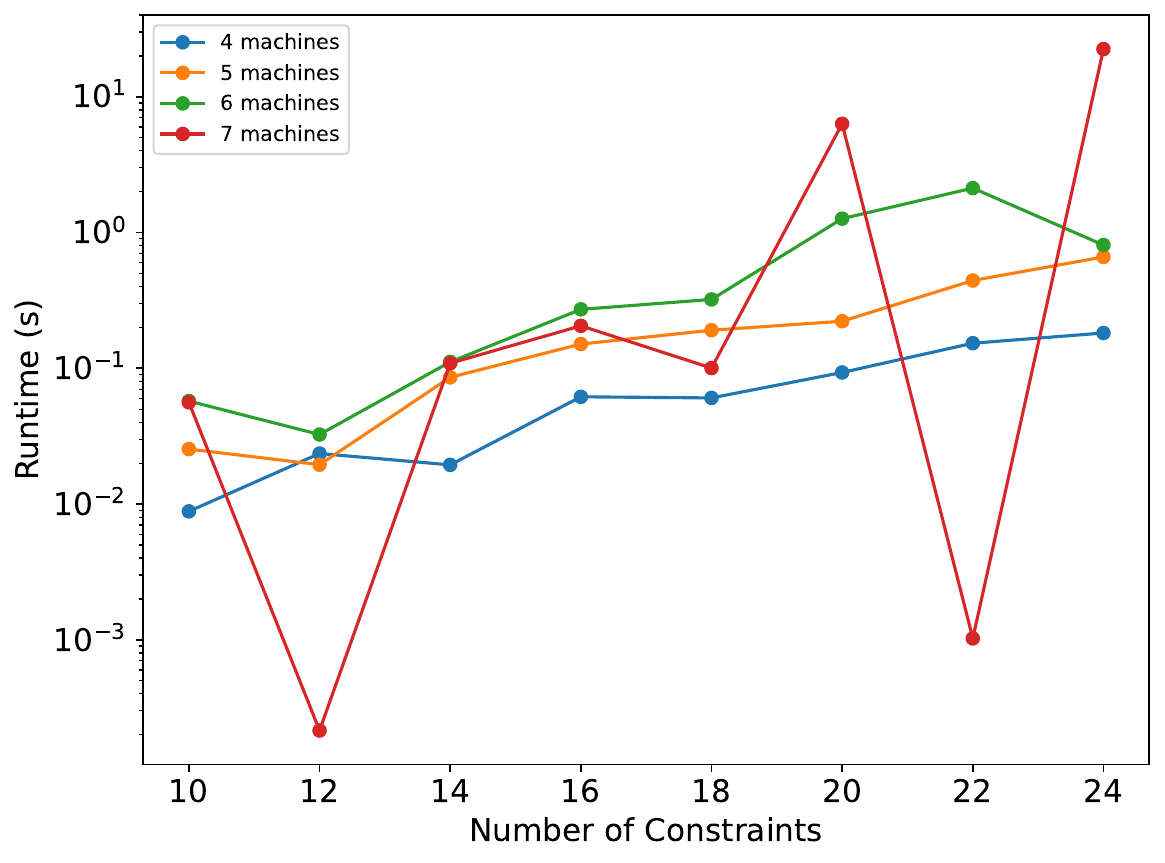}
        \caption{Runtime against number of constraints $n_r$, for different number of machines. There are $5$ tasks.}
        \label{fig: scaling constraints base}
    \end{subfigure}
    \caption{Scaling of the mean runtime of the normal algorithm with respect to different parameters. There are $10$ samples per point and $\tau=100$.}
    \label{fig:scaling base}
\end{figure*}

\begin{figure*}
    \centering
    \begin{subfigure}[b]{0.32\linewidth}
        \centering
        \includegraphics[width=\linewidth]{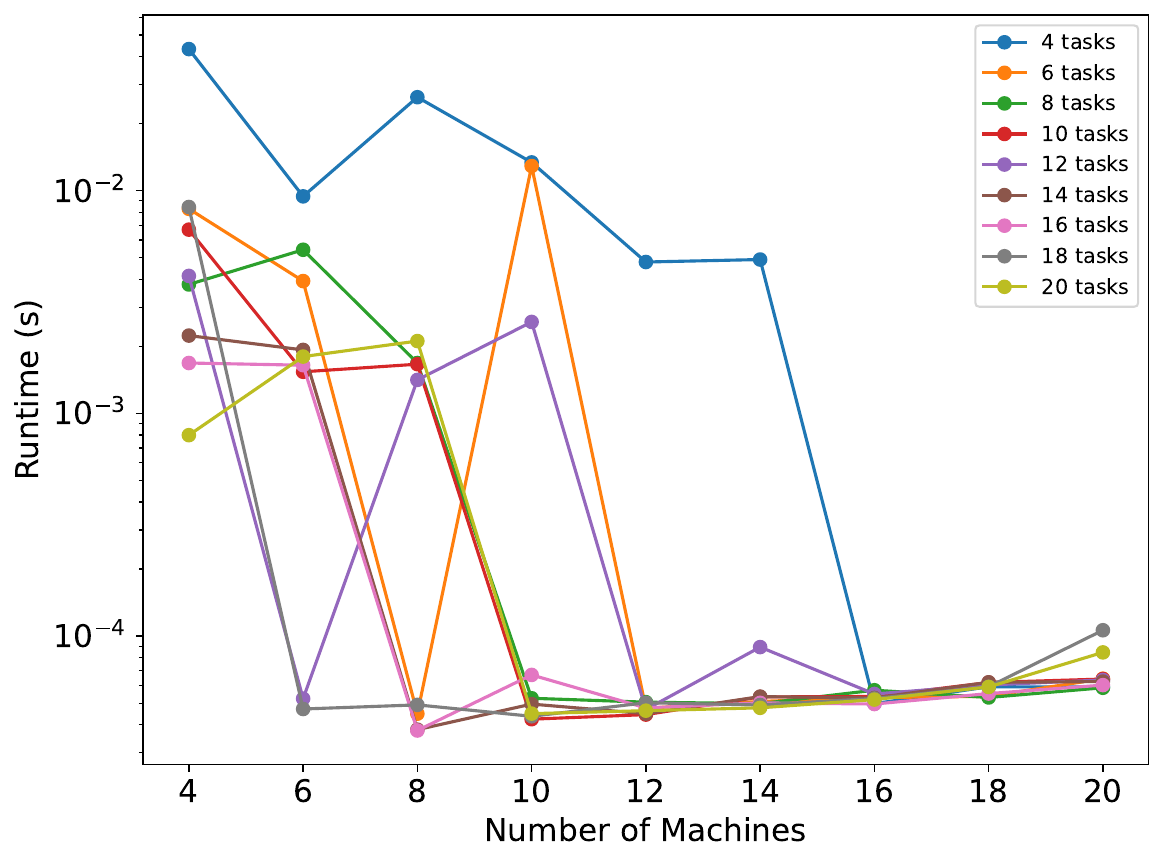}
        \caption{Runtime against number of machines $m$, for different number of tasks. There are $25$ constraints.}
        \label{fig: scaling mach iter}
    \end{subfigure}
    \hfill
    \begin{subfigure}[b]{0.32\linewidth}
        \centering
        \includegraphics[width=\linewidth]{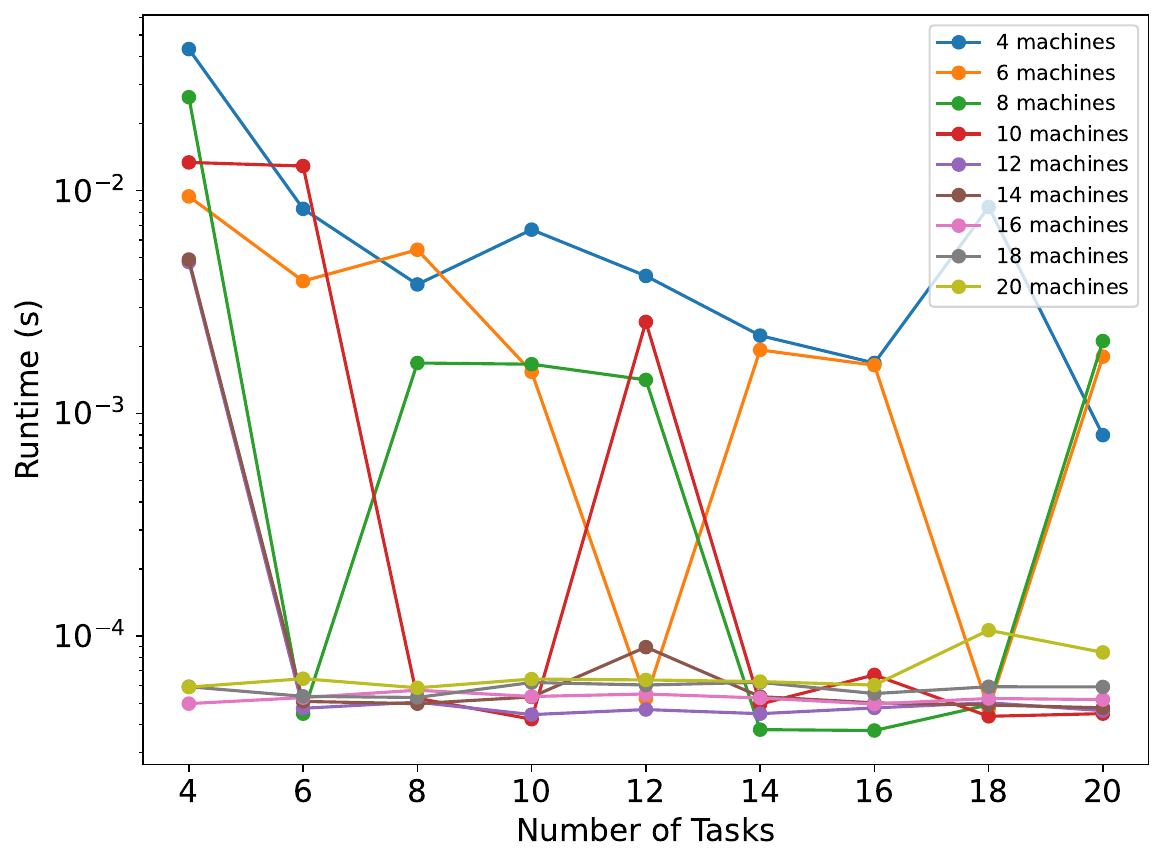}
        \caption{Runtime against number of tasks $P$, for different number of machines. There are $25$ constraints.}
        \label{fig: scaling tasks iter}
    \end{subfigure}
    \hfill
    \begin{subfigure}[b]{0.32\linewidth}
        \centering
        \includegraphics[width=\linewidth]{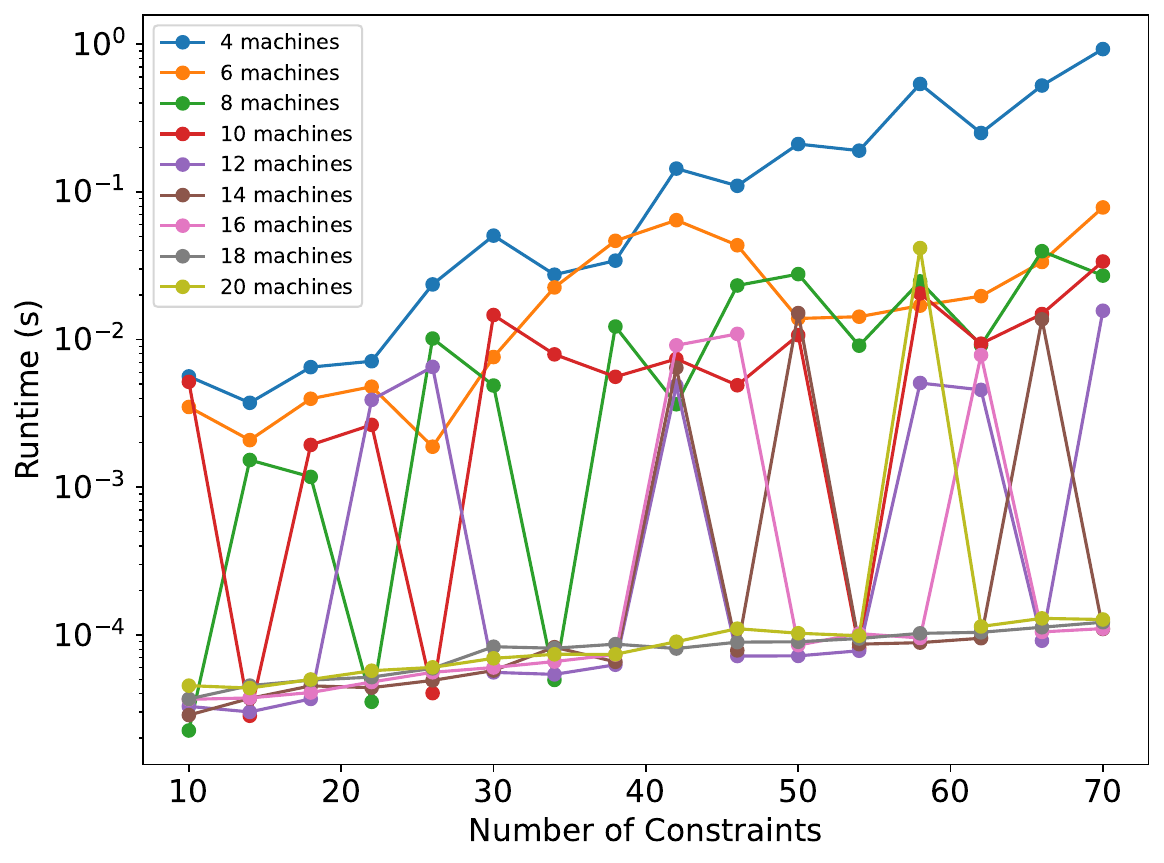}
        \caption{Runtime against number of constraints $n_r$, for different number of machines. There are $5$ tasks.}
        \label{fig: scaling constraints iter}
    \end{subfigure}
    \caption{Scaling of the mean runtime of the iterative algorithm with respect to different parameters. There are $10$ samples per point and $\tau=100$.}
    \label{fig:scaling iter}
\end{figure*}

First, the runtime of normal and iterative algorithms is tested in Figs.~\ref{fig:scaling base} and \ref{fig:scaling iter}. These tests compare the mean runtime of each algorithm to obtain the solution, with $10$ samples per point. Both figures show that the iterative algorithm is much faster than the normal algorithm, even with more calls to the solver. In both cases, Figs.~\ref{fig: scaling constraints base} and \ref{fig: scaling constraints iter} show that the number of constraints increases the runtime. In Figs.~\ref{fig: scaling mach base} and \ref{fig: scaling tasks base}, normal algorithm does not show a clear behavior for the number of tasks and machines. However, this is because of the limited size of experiments that can be performed due to the high memory scaling of the normal algorithm.

 In the iterative case, Figs.~\ref{fig: scaling mach iter} and \ref{fig: scaling tasks iter} show that the runtime decreases with the number of machines for a fixed number of tasks and constraints. This is not because of algorithm speed. This behavior is due to the fact that with more machines it is more simple that the constraints do not restrict the solution space enough to require all the constraints.

\begin{figure*}
    \centering
    \begin{subfigure}[b]{0.32\linewidth}
        \centering
        \includegraphics[width=\linewidth]{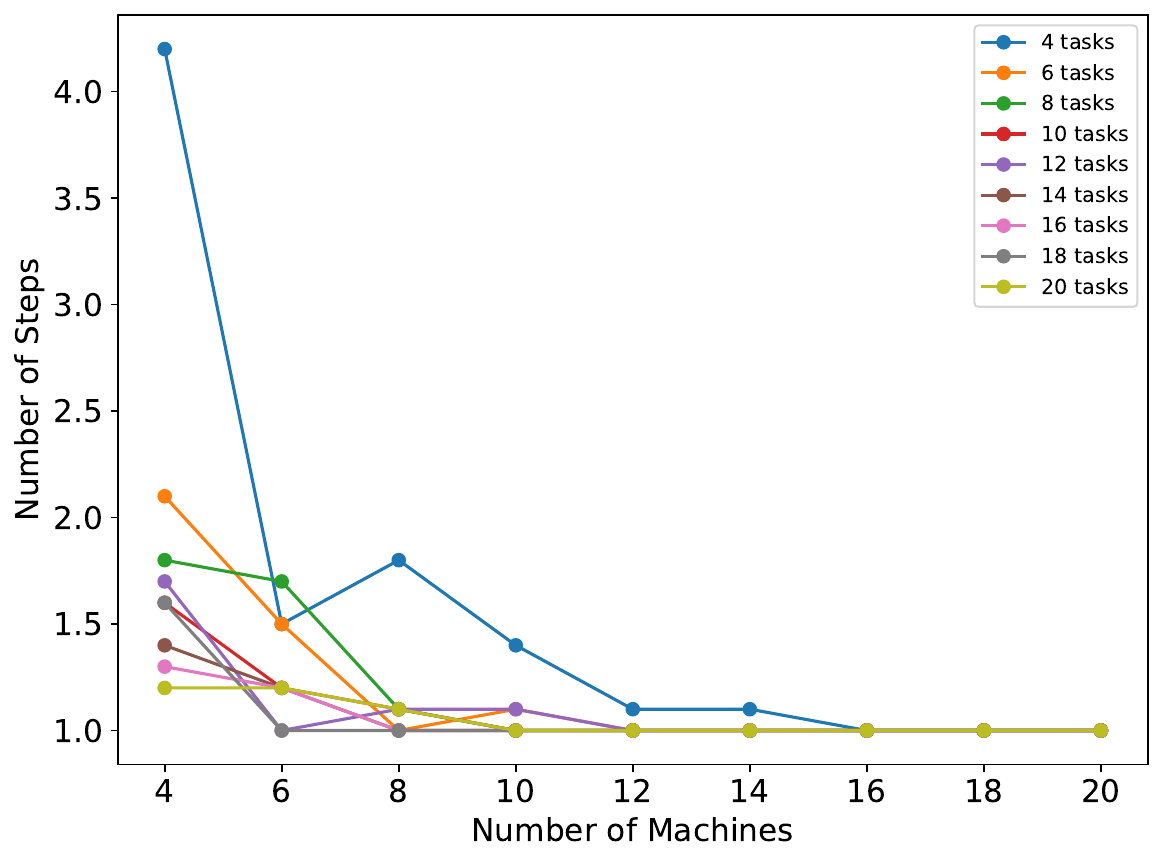}
        \caption{Steps against number of machines $m$, for different number of tasks. There are $25$ constraints.}
        \label{fig: steps mach iter}
    \end{subfigure}
    \hfill
    \begin{subfigure}[b]{0.32\linewidth}
        \centering
        \includegraphics[width=\linewidth]{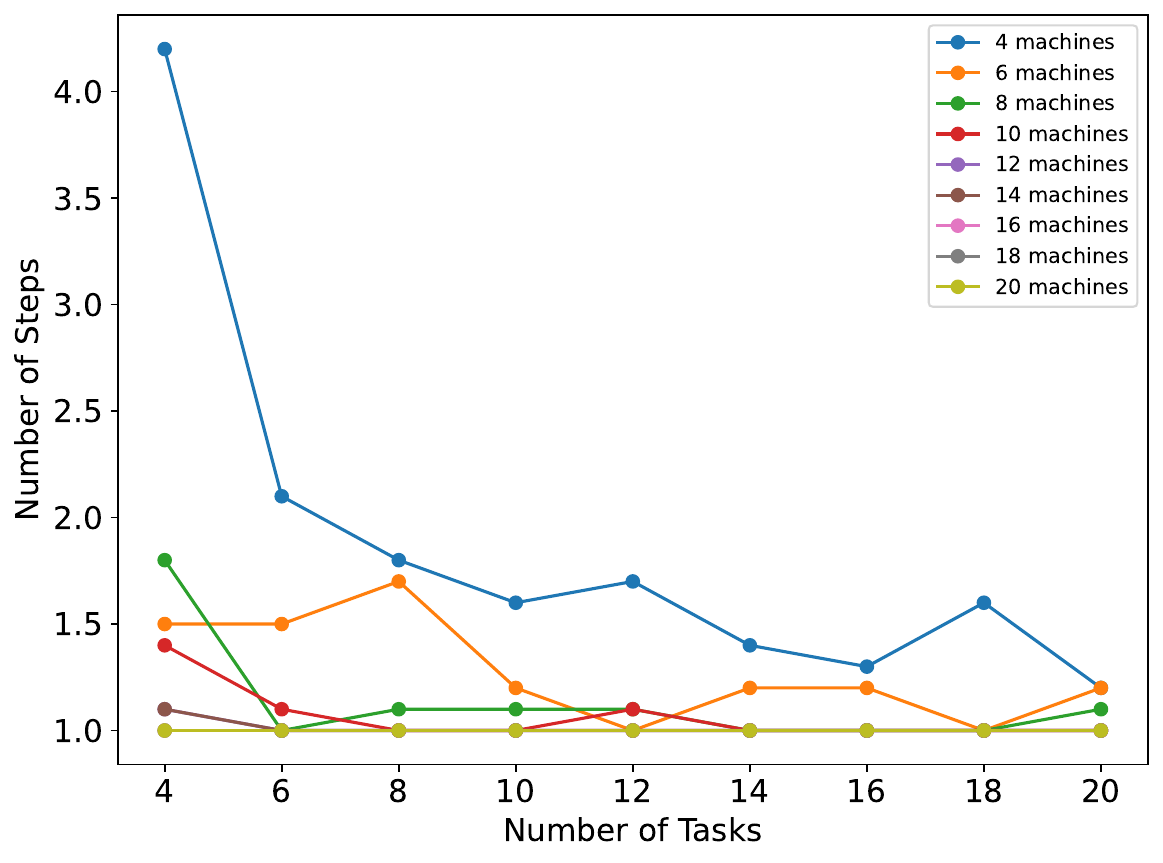}
        \caption{Steps against number of tasks $P$, for different number of machines. There are $25$ constraints.}
        \label{fig: steps tasks iter}
    \end{subfigure}
    \hfill
    \begin{subfigure}[b]{0.32\linewidth}
        \centering
        \includegraphics[width=\linewidth]{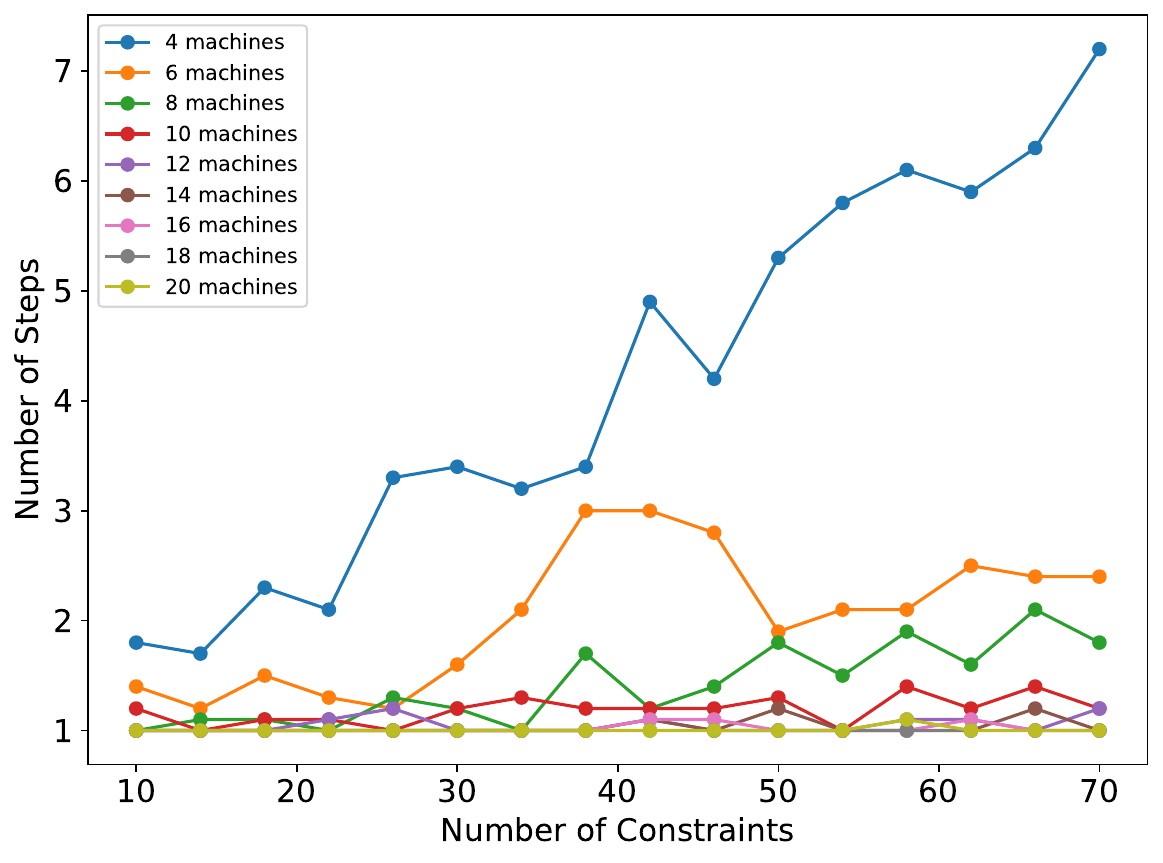}
        \caption{Steps against number of constraints $n_r$, for different number of machines. There are $5$ tasks.}
        \label{fig: steps constraints iter}
    \end{subfigure}
    \caption{Scaling of the mean number of required steps of the iterative algorithm with respect to different parameters. There are $10$ samples per point and $\tau=100$.}
    \label{fig:steps iter}
\end{figure*}

Another interesting experiment is to determine the scaling of the number of required steps in the iterative algorithm to obtain the correct solution. Figs.~\ref{fig:steps iter} show the number of steps required for the previous experiments. Figs.~\ref{fig: steps mach iter} and \ref{fig: steps tasks iter} show that the number of steps decreases with the number of machines and tasks, due to the same reasons presented before. However, Fig.~\ref{fig: steps constraints iter} shows that the number of steps increases with the number of constraints, faster for a smaller number of machines. This is expected because fewer machines imply more situations where all the constraints are relevant.

\begin{figure}
    \centering
        \includegraphics[width=0.7\linewidth]{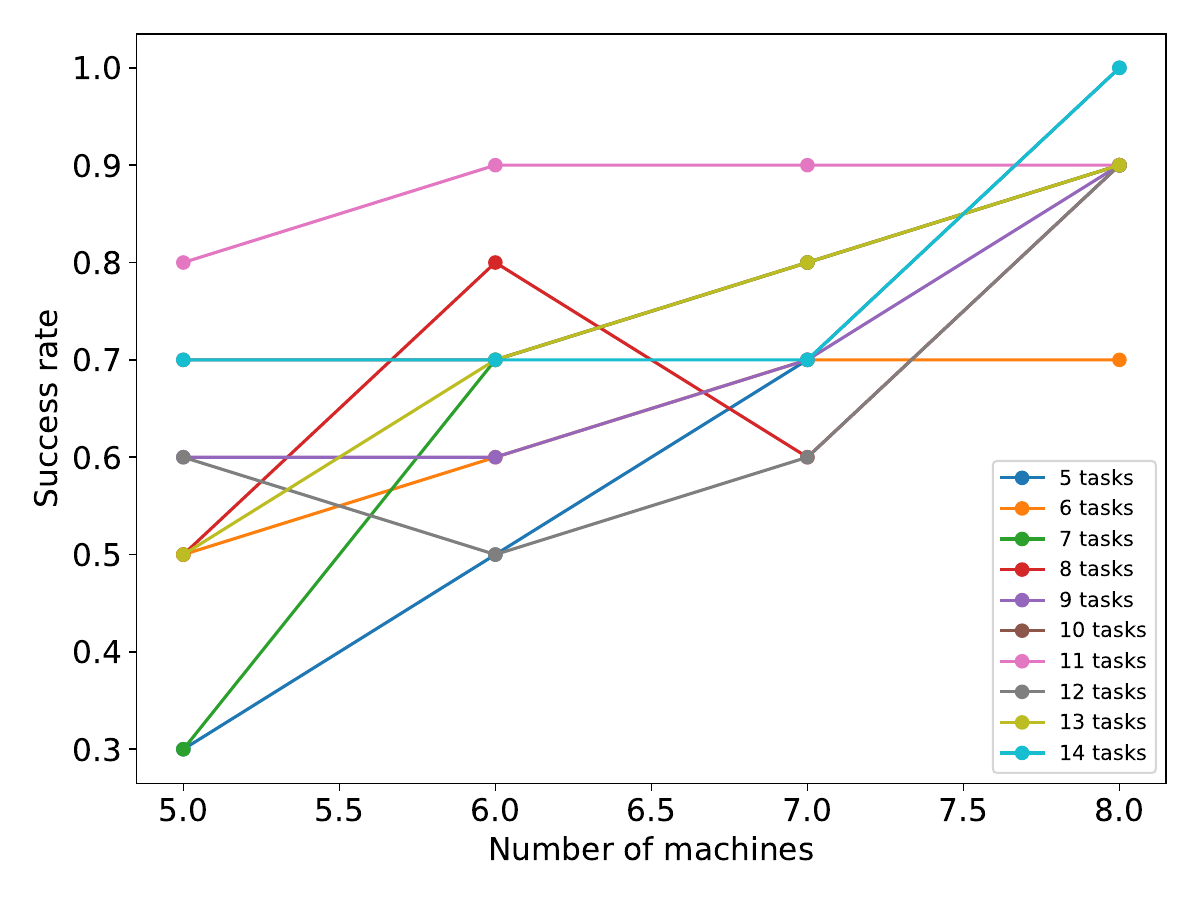}
    \caption{Mean success rate of the genetic algorithm against the number of machines $m$ for different number of tasks. There are $30$ constraints, $10$ samples per point and $\tau=100$.}
    \label{fig:genetic}
\end{figure}

Finally, the performance of the genetic algorithm is tested. Fig.~\ref{fig:genetic} shows the success rate of the genetic algorithm to determine a valid combination against the number of machines, for different number of tasks. The number of individuals and generations is set to $30$, the maximum number of constraints per individual is set to $10$, the number of tasks per individual is set to $4$, the proportion of parents is set to $1/3$ and the number of crosses and the number of mutations per individual and crossover are set to $5$. These parameters are set to have individuals that are more easily executable than the main problem. The larger number of generations and individuals does not appear to improve the performance of the algorithm in the tested instances. The result of the experiments is that the success rate does not seem to depend on the number of tasks, but it increases with the number of machines, probably due to the reasons presented before. However, the success rate is too low, taking into account that these instances are solvable with the iterative method exactly. This indicates that the genetic algorithm does not work properly with this type of problem, at least not implemented in the way presented.

\section{Conclusions}\label{sec: conclusions}
This work developed a novel quantum-inspired algorithm for combinatorial optimization and applied it to the industrial case of machine task distribution with directed constraints, minimizing the total execution cost. Even with the improvements presented, these algorithms have the limitation of exponential scaling in the worst-case scenario. This situation is not highly probable, but it may be a problem for these kinds of algorithms in production environments. This method has been tested with several instances and problems of various sizes. The iterative algorithm succeeds in improving the scalability of the normal algorithm, and the tests show that in highly random constraint situations, similar to those extracted from historical data, most of the constraints can be neglected, even more in larger systems. However, the genetic algorithm does not perform well, being unable to obtain compatible solutions. An interesting future line of research is the study of better heuristic and approximated methods for the search for optimal sub-problems to solve the problem. Moreover, its combination with genetic algorithms indicates a possible synergy between both technologies, but it needs to improve the mechanism of combination.

The method can also be extended in future work for other types of bidirectional constraints or restrictions where only certain combinations are possible. This can provide a wider range of applications for this kind of algorithm. Another future study would be to further analyze the algorithm to improve its computation, for example by using Matrix Product State (MPS/TT) compression~\cite{Compress, PEPS}, when constraints are applied or by condensing the constraints in a better way, with techniques such as those presented in~\cite{cons_train}. The general algorithm could also be tested for different significant problems, such as the Traveling Salesman problem or the Job Shop Scheduling problem.

\nocite{*}

\bibliography{apssamp}% Produces the bibliography via BibTeX.

@PREAMBLE{
 "\providecommand{\noopsort}[1]{}" 
 # "\providecommand{\singleletter}[1]{#1}%" 
}

@Article{FlowShop1,
author={Liang, Zhongyuan
and Zhong, Peisi
and Liu, Mei
and Zhang, Chao
and Zhang, Zhenyu},
title={A computational efficient optimization of flow shop scheduling problems},
journal={Scientific Reports},
year={2022},
month={Jan},
day={17},
volume={12},
number={1},
pages={845},
issn={2045-2322},
doi={10.1038/s41598-022-04887-8},
url={https://doi.org/10.1038/s41598-022-04887-8}
}

@article{FlowShop2,
title = {A Simple Model to Optimize General Flow-Shop Scheduling Problems With Known Break Down Time And Weights Of Jobs},
journal = {Procedia Engineering},
volume = {38},
pages = {191-196},
year = {2012},
note = {INTERNATIONAL CONFERENCE ON MODELLING OPTIMIZATION AND COMPUTING},
issn = {1877-7058},
doi = {https://doi.org/10.1016/j.proeng.2012.06.026},
url = {https://www.sciencedirect.com/science/article/pii/S187770581201939X},
author = {Baskar A and Anthony Xavior M}
}

@article{FlowShop3, title={Learning to Optimize Permutation Flow Shop Scheduling via Graph-Based Imitation Learning}, volume={38}, url={https://ojs.aaai.org/index.php/AAAI/article/view/29998}, DOI={10.1609/aaai.v38i18.29998}, abstractNote={The permutation flow shop scheduling (PFSS), aiming at finding the optimal permutation of jobs, is widely used in manufacturing systems. When solving large-scale PFSS problems, traditional optimization algorithms such as heuristics could hardly meet the demands of both solution accuracy and computational efficiency, thus learning-based methods have recently garnered more attention. Some work attempts to solve the problems by reinforcement learning methods, which suffer from slow convergence issues during training and are still not accurate enough regarding the solutions. To that end, we propose to train the model via expert-driven imitation learning, which accelerates convergence more stably and accurately. Moreover, in order to extract better feature representations of input jobs, we incorporate the graph structure as the encoder. The extensive experiments reveal that our proposed model obtains significant promotion and presents excellent generalizability in large-scale problems with up to 1000 jobs. Compared to the state-of-the-art reinforcement learning method, our model’s network parameters are reduced to only 37% of theirs, and the solution gap of our model towards the expert solutions decreases from 6.8% to 1.3% on average. The code is available at: https://github.com/longkangli/PFSS-IL.}, number={18}, journal={Proceedings of the AAAI Conference on Artificial Intelligence}, author={Li, Longkang and Liang, Siyuan and Zhu, Zihao and Ding, Chris and Zha, Hongyuan and Wu, Baoyuan}, year={2024}, month={Mar.}, pages={20185-20193} }

@INPROCEEDINGS{Genetic,
  author={Pelikan, M. and Goldberg, D.E. and Lobo, F.},
  booktitle={Proceedings of the 2000 American Control Conference. ACC (IEEE Cat. No.00CH36334)}, 
  title={A survey of optimization by building and using probabilistic models}, 
  year={2000},
  volume={5},
  number={},
  pages={3289-3293 vol.5},
  doi={10.1109/ACC.2000.879173}}

@InProceedings{Enjambre,
author="Karaboga, Dervis
and Basturk, Bahriye",
editor="Melin, Patricia
and Castillo, Oscar
and Aguilar, Luis T.
and Kacprzyk, Janusz
and Pedrycz, Witold",
title="Artificial Bee Colony (ABC) Optimization Algorithm for Solving Constrained Optimization Problems",
booktitle="Foundations of Fuzzy Logic and Soft Computing",
year="2007",
publisher="Springer Berlin Heidelberg",
address="Berlin, Heidelberg",
pages="789--798",
isbn="978-3-540-72950-1"
}

@misc{QAOA,
      title={A Quantum Approximate Optimization Algorithm}, 
      author={Edward Farhi and Jeffrey Goldstone and Sam Gutmann},
      year={2014},
      eprint={1411.4028},
      archivePrefix={arXiv},
      primaryClass={quant-ph}
}

@article{VQE,
title = {The Variational Quantum Eigensolver: A review of methods and best practices},
journal = {Physics Reports},
volume = {986},
pages = {1-128},
year = {2022},
note = {The Variational Quantum Eigensolver: a review of methods and best practices},
issn = {0370-1573},
doi = {https://doi.org/10.1016/j.physrep.2022.08.003},
url = {https://www.sciencedirect.com/science/article/pii/S0370157322003118},
author = {Jules Tilly and Hongxiang Chen and Shuxiang Cao and Dario Picozzi and Kanav Setia and Ying Li and Edward Grant and Leonard Wossnig and Ivan Rungger and George H. Booth and Jonathan Tennyson}
}

@article{Annealing,
  title = {Quantum Annealing for Constrained Optimization},
  author = {Hen, Itay and Spedalieri, Federico M.},
  journal = {Phys. Rev. Appl.},
  volume = {5},
  issue = {3},
  pages = {034007},
  numpages = {7},
  year = {2016},
  month = {Mar},
  publisher = {American Physical Society},
  doi = {10.1103/PhysRevApplied.5.034007},
  url = {https://link.aps.org/doi/10.1103/PhysRevApplied.5.034007}
}

@article{Digital,
   title={Physics-Inspired Optimization for Quadratic Unconstrained Problems Using a Digital Annealer},
   volume={7},
   ISSN={2296-424X},
   url={http://dx.doi.org/10.3389/fphy.2019.00048},
   DOI={10.3389/fphy.2019.00048},
   journal={Frontiers in Physics},
   publisher={Frontiers Media SA},
   author={Aramon, Maliheh and Rosenberg, Gili and Valiante, Elisabetta and Miyazawa, Toshiyuki and Tamura, Hirotaka and Katzgraber, Helmut G.},
   year={2019},
   month=apr }

@misc{Tensor,
      title={Tensor Networks in a Nutshell}, 
      author={Jacob Biamonte and Ville Bergholm},
      year={2017},
      eprint={1708.00006},
      archivePrefix={arXiv},
      primaryClass={quant-ph}
}

@inproceedings{TTOpt,
 author = {Sozykin, Konstantin and Chertkov, Andrei and Schutski, Roman and Phan, Anh-Huy and CICHOCKI, Andrzej S and Oseledets, Ivan},
 booktitle = {Advances in Neural Information Processing Systems},
 editor = {S. Koyejo and S. Mohamed and A. Agarwal and D. Belgrave and K. Cho and A. Oh},
 pages = {26052--26065},
 publisher = {Curran Associates, Inc.},
 title = {TTOpt: A Maximum Volume Quantized Tensor Train-based Optimization and its Application to Reinforcement Learning},
 url = {https://proceedings.neurips.cc/paper_files/paper/2022/file/a730abbcd6cf4a371ca9545db5922442-Paper-Conference.pdf},
 volume = {35},
 year = {2022}
}

@Article{GEO,
author={Alcazar, Javier
and Ghazi Vakili, Mohammad
and Kalayci, Can B.
and Perdomo-Ortiz, Alejandro},
title={Enhancing combinatorial optimization with classical and quantum generative models},
journal={Nature Communications},
year={2024},
month={Mar},
day={29},
volume={15},
number={1},
pages={2761},
issn={2041-1723},
doi={10.1038/s41467-024-46959-5},
url={https://doi.org/10.1038/s41467-024-46959-5}
}

@ARTICLE{Combin,
  
AUTHOR={Hao, Tianyi and Huang, Xuxin and Jia, Chunjing and Peng, Cheng},   
	 
TITLE={A Quantum-Inspired Tensor Network Algorithm for Constrained Combinatorial Optimization Problems},      
	
JOURNAL={Frontiers in Physics},      
	
VOLUME={10},           
	
YEAR={2022},      
	  
URL={https://www.frontiersin.org/articles/10.3389/fphy.2022.906590},       
	
DOI={10.3389/fphy.2022.906590},      
	
ISSN={2296-424X}
}

@misc{QUDO,
      title={Polynomial-time Solver of Tridiagonal QUBO, QUDO and Tensor QUDO problems with Tensor Networks}, 
      author={Alejandro Mata Ali and Iñigo Perez Delgado and Marina Ristol Roura and Aitor Moreno Fdez. de Leceta},
      year={2025},
      eprint={2309.10509},
      archivePrefix={arXiv},
      primaryClass={quant-ph},
      url={https://arxiv.org/abs/2309.10509}, 
}

@misc{Grover,
      title={A fast quantum mechanical algorithm for database search}, 
      author={Lov K. Grover},
      year={1996},
      eprint={quant-ph/9605043},
      archivePrefix={arXiv},
      primaryClass={quant-ph}
}

@misc{TN_lib,
      title={TensorNetwork: A Library for Physics and Machine Learning}, 
      author={Chase Roberts and Ashley Milsted and Martin Ganahl and Adam Zalcman and Bruce Fontaine and Yijian Zou and Jack Hidary and Guifre Vidal and Stefan Leichenauer},
      year={2019},
      eprint={1905.01330},
      archivePrefix={arXiv},
      primaryClass={physics.comp-ph}
}

@article{Compress,
   title={Quantum compression of tensor network states},
   volume={22},
   ISSN={1367-2630},
   url={http://dx.doi.org/10.1088/1367-2630/ab7a34},
   DOI={10.1088/1367-2630/ab7a34},
   number={4},
   journal={New Journal of Physics},
   publisher={IOP Publishing},
   author={Bai, Ge and Yang, Yuxiang and Chiribella, Giulio},
   year={2020},
   month=apr, pages={043015} }

@article{PEPS,
   title={Unifying projected entangled pair state contractions},
   volume={16},
   ISSN={1367-2630},
   url={http://dx.doi.org/10.1088/1367-2630/16/3/033014},
   DOI={10.1088/1367-2630/16/3/033014},
   number={3},
   journal={New Journal of Physics},
   publisher={IOP Publishing},
   author={Lubasch, Michael and Cirac, J Ignacio and Bañuls, Mari-Carmen},
   year={2014},
   month=mar, pages={033014} }

@misc{melocoton,
      title={Explicit Solution Equation for Every Combinatorial Problem via Tensor Networks: MeLoCoToN}, 
      author={Alejandro Mata Ali},
      year={2025},
      eprint={2502.05981},
      archivePrefix={arXiv},
      primaryClass={cs.ET},
      url={https://arxiv.org/abs/2502.05981}, 
}

@inproceedings{bottleneck_RCPS,
   title={Bottleneck Identification in Resource-Constrained Project Scheduling via Constraint Relaxation},
   url={http://dx.doi.org/10.5220/0013253700003893},
   DOI={10.5220/0013253700003893},
   booktitle={Proceedings of the 14th International Conference on Operations Research and Enterprise Systems},
   publisher={SCITEPRESS - Science and Technology Publications},
   author={Nedbálek, Lukáš and Novák, Antonín},
   year={2025},
   pages={340–347} }

@article{Mixed_JobShop,
title = {Mixed Integer Programming models for job shop scheduling: A computational analysis},
journal = {Computers \& Operations Research},
volume = {73},
pages = {165-173},
year = {2016},
issn = {0305-0548},
doi = {https://doi.org/10.1016/j.cor.2016.04.006},
url = {https://www.sciencedirect.com/science/article/pii/S0305054816300764},
author = {Wen-Yang Ku and J. Christopher Beck}
}

@article{extension_RCPSP,
title = {Extensions of the resource-constrained project scheduling problem},
journal = {Automation in Construction},
volume = {153},
pages = {104958},
year = {2023},
issn = {0926-5805},
doi = {https://doi.org/10.1016/j.autcon.2023.104958},
url = {https://www.sciencedirect.com/science/article/pii/S0926580523002182},
author = {Hongyan Ding and Cunbo Zhuang and Jianhua Liu}
}

@article{NP_Complete,
title = {NP-complete scheduling problems},
journal = {Journal of Computer and System Sciences},
volume = {10},
number = {3},
pages = {384-393},
year = {1975},
issn = {0022-0000},
doi = {https://doi.org/10.1016/S0022-0000(75)80008-0},
url = {https://www.sciencedirect.com/science/article/pii/S0022000075800080},
author = {J.D. Ullman}
}

@InProceedings{Precedence,
author="van Bevern, Ren{\'e}
and Bredereck, Robert
and Bulteau, Laurent
and Komusiewicz, Christian
and Talmon, Nimrod
and Woeginger, Gerhard J.",
editor="Kochetov, Yury
and Khachay, Michael
and Beresnev, Vladimir
and Nurminski, Evgeni
and Pardalos, Panos",
title="Precedence-Constrained Scheduling Problems Parameterized by Partial Order Width",
booktitle="Discrete Optimization and Operations Research",
year="2016",
publisher="Springer International Publishing",
address="Cham",
pages="105--120",
isbn="978-3-319-44914-2"
}

@article{Resource_complex,
title = {Scheduling subject to resource constraints: classification and complexity},
journal = {Discrete Applied Mathematics},
volume = {5},
number = {1},
pages = {11-24},
year = {1983},
issn = {0166-218X},
doi = {https://doi.org/10.1016/0166-218X(83)90012-4},
url = {https://www.sciencedirect.com/science/article/pii/0166218X83900124},
author = {J. Blazewicz and J.K. Lenstra and A.H.G.Rinnooy Kan}
}

@inproceedings{Complexity_lands,
author = {Ganian, Robert and Hamm, Thekla and Mescoff, Guillaume},
title = {The complexity landscape of resource-constrained scheduling},
year = {2021},
isbn = {9780999241165},
booktitle = {Proceedings of the Twenty-Ninth International Joint Conference on Artificial Intelligence},
articleno = {241},
numpages = {7},
location = {Yokohama, Yokohama, Japan},
series = {IJCAI'20}
}

@INPROCEEDINGS{genetic_rcps,
  author={Dridi, Olfa and Krichen, Saoussen and Guitouni, Adel},
  booktitle={2013 5th International Conference on Modeling, Simulation and Applied Optimization (ICMSAO)}, 
  title={Solving resource-constrained project scheduling problem by a genetic local search approach}, 
  year={2013},
  volume={},
  number={},
  pages={1-5},
  doi={10.1109/ICMSAO.2013.6552544}}

@article{genetic_flexible,
title = {A genetic algorithm for the Flexible Job-shop Scheduling Problem},
journal = {Computers \& Operations Research},
volume = {35},
number = {10},
pages = {3202-3212},
year = {2008},
note = {Part Special Issue: Search-based Software Engineering},
issn = {0305-0548},
doi = {https://doi.org/10.1016/j.cor.2007.02.014},
url = {https://www.sciencedirect.com/science/article/pii/S0305054807000524},
author = {F. Pezzella and G. Morganti and G. Ciaschetti}
}

@article{genetic_job_shop,
title = {A genetic algorithm for the job shop problem},
journal = {Computers \& Operations Research},
volume = {22},
number = {1},
pages = {15-24},
year = {1995},
note = {Genetic Algorithms},
issn = {0305-0548},
doi = {https://doi.org/10.1016/0305-0548(93)E0015-L},
url = {https://www.sciencedirect.com/science/article/pii/0305054893E0015L},
author = {Federico {Della Croce} and Roberto Tadei and Giuseppe Volta}
}

@article{genetic_flow_shop,
title = {Genetic algorithms for flowshop scheduling problems},
journal = {Computers \& Industrial Engineering},
volume = {30},
number = {4},
pages = {1061-1071},
year = {1996},
issn = {0360-8352},
doi = {https://doi.org/10.1016/0360-8352(96)00053-8},
url = {https://www.sciencedirect.com/science/article/pii/0360835296000538},
author = {Tadahiko Murata and Hisao Ishibuchi and Hideo Tanaka}
}

@article{integer_flow,
title = {A new integer programming formulation for the permutation flowshop problem},
journal = {European Journal of Operational Research},
volume = {40},
number = {1},
pages = {90-98},
year = {1989},
issn = {0377-2217},
doi = {https://doi.org/10.1016/0377-2217(89)90276-2},
url = {https://www.sciencedirect.com/science/article/pii/0377221789902762},
author = {A.M. Frieze and J. Yadegar}
}

@article{tabu_search_job_shop,
title = {A tabu search algorithm with a new neighborhood structure for the job shop scheduling problem},
journal = {Computers \& Operations Research},
volume = {34},
number = {11},
pages = {3229-3242},
year = {2007},
issn = {0305-0548},
doi = {https://doi.org/10.1016/j.cor.2005.12.002},
url = {https://www.sciencedirect.com/science/article/pii/S0305054805003989},
author = {ChaoYong Zhang and PeiGen Li and ZaiLin Guan and YunQing Rao}
}

@Article{Tabu_job_shop_2,
author={Dell'Amico, Mauro
and Trubian, Marco},
title={Applying tabu search to the job-shop scheduling problem},
journal={Annals of Operations Research},
year={1993},
month={Sep},
day={01},
volume={41},
number={3},
pages={231-252},
issn={1572-9338},
doi={10.1007/BF02023076},
url={https://doi.org/10.1007/BF02023076}
}

@Article{Q_annealing_rcps,
author={P{\'e}rez Armas, Luis Fernando
and Creemers, Stefan
and Deleplanque, Samuel},
title={Solving the resource constrained project scheduling problem with quantum annealing},
journal={Scientific Reports},
year={2024},
month={Jul},
day={22},
volume={14},
number={1},
pages={16784},
issn={2045-2322},
doi={10.1038/s41598-024-67168-6},
url={https://doi.org/10.1038/s41598-024-67168-6}
}

@article{Q_annealing_flexible,
title = {Multi-objective Quantum Annealing approach for solving flexible job shop scheduling in manufacturing},
journal = {Journal of Manufacturing Systems},
volume = {72},
pages = {142-153},
year = {2024},
issn = {0278-6125},
doi = {https://doi.org/10.1016/j.jmsy.2023.11.015},
url = {https://www.sciencedirect.com/science/article/pii/S027861252300242X},
author = {Philipp Schworm and Xiangqian Wu and Matthias Klar and Moritz Glatt and Jan C. Aurich}
}

@article{qaoa_jssp,
title = {Application of quantum approximate optimization algorithm to job shop scheduling problem},
journal = {European Journal of Operational Research},
volume = {310},
number = {2},
pages = {518-528},
year = {2023},
issn = {0377-2217},
doi = {https://doi.org/10.1016/j.ejor.2023.03.013},
url = {https://www.sciencedirect.com/science/article/pii/S0377221723002072},
author = {Krzysztof Kurowski and Tomasz Pecyna and Mateusz Slysz and Rafał Różycki and Grzegorz Waligóra and Jan Wȩglarz}
}

@Article{Q_Annealer_JSSP_Dwave,
author={Carugno, Costantino
and Ferrari Dacrema, Maurizio
and Cremonesi, Paolo},
title={Evaluating the job shop scheduling problem on a D-wave quantum annealer},
journal={Scientific Reports},
year={2022},
month={Apr},
day={21},
volume={12},
number={1},
pages={6539},
issn={2045-2322},
doi={10.1038/s41598-022-10169-0},
url={https://doi.org/10.1038/s41598-022-10169-0}
}

@Article{q_annealer_flexible_2,
author={Schworm, Philipp
and Wu, Xiangqian
and Glatt, Moritz
and Aurich, Jan C.},
title={Solving flexible job shop scheduling problems in manufacturing with Quantum Annealing},
journal={Production Engineering},
year={2023},
month={Feb},
day={01},
volume={17},
number={1},
pages={105-115},
issn={1863-7353},
doi={10.1007/s11740-022-01145-8},
url={https://doi.org/10.1007/s11740-022-01145-8}
}

@INPROCEEDINGS{iterative_job_shop,
  author={Ishigaki, Aya and Takaki, Shun},
  booktitle={2017 6th IIAI International Congress on Advanced Applied Informatics (IIAI-AAI)}, 
  title={Iterated Local Search Algorithm for Flexible Job Shop Scheduling}, 
  year={2017},
  volume={},
  number={},
  pages={947-952},
  doi={10.1109/IIAI-AAI.2017.126}}

@article{iterative_flow_shop,
title = {An effective iterated local search algorithm for the distributed no-wait flowshop scheduling problem},
journal = {Engineering Applications of Artificial Intelligence},
volume = {120},
pages = {105921},
year = {2023},
issn = {0952-1976},
doi = {https://doi.org/10.1016/j.engappai.2023.105921},
url = {https://www.sciencedirect.com/science/article/pii/S0952197623001057},
author = {Mustafa Avci}
}

@Article{cons_train,
	title={{Cons-training tensor networks: Embedding and optimization over discrete linear constraints}},
	author={Javier Lopez-Piqueres and Jing Chen},
	journal={SciPost Phys.},
	volume={18},
	pages={192},
	year={2025},
	publisher={SciPost},
	doi={10.21468/SciPostPhys.18.6.192},
	url={https://scipost.org/10.21468/SciPostPhys.18.6.192},
}

\end{document}